%% file: Benedict_language_edition.tex
\documentclass{aa} 
\usepackage{color}
\usepackage{graphicx}
\usepackage{txfonts}
\usepackage{natbib,twoopt}
\usepackage[breaklinks=true]{hyperref}
\usepackage{lscape}
\usepackage{colortbl}
\usepackage{multirow}
\usepackage{caption} 
\hypersetup{colorlinks=true,linkcolor=blue,citecolor=blue,urlcolor=cyan}

\newcommand\T{\rule{0pt}{3.0ex}}       
\newcommand\B{\rule[-1.6ex]{0pt}{0pt}} 

\bibpunct{(}{)}{;}{a}{}{,} 
\makeatletter
\newcommandtwoopt{\citeads}[3][][]{\href{http://adsabs.harvard.edu/abs/#3}%
{\def\hyper@linkstart##1##2{}%
\let\hyper@linkend\@empty\citealp[#1][#2]{#3}}}
\newcommandtwoopt{\citepads}[3][][]{\href{http://adsabs.harvard.edu/abs/#3}%
{\def\hyper@linkstart##1##2{}%
\let\hyper@linkend\@empty\citep[#1][#2]{#3}}}
\newcommandtwoopt{\citetads}[3][][]{\href{http://adsabs.harvard.edu/abs/#3}%
{\def\hyper@linkstart##1##2{}%
\let\hyper@linkend\@empty\citet[#1][#2]{#3}}}
\newcommandtwoopt{\citeyearads}[3][][]%
{\href{http://adsabs.harvard.edu/abs/#3}
{\def\hyper@linkstart##1##2{}%
\let\hyper@linkend\@empty\citeyear[#1][#2]{#3}}}
\makeatother

\definecolor{forestgreen}{rgb}{0.11,0.54,0.15}
\definecolor{purple}{rgb}{0.62,0.10,0.96}

\begin{document}

   \title{Observational calibration of the projection factor of Cepheids}

   \subtitle{II. Application to nine Cepheids with HST/FGS parallax measurements\thanks{Based on observations carried out with ESO facilities at Paranal Observatory under program 093.D-0316, 094.D-0773 and 094.D-0584}}

   \author{J. Breitfelder\inst{1, 2}
          \and
          A. M\'{e}rand\inst{1}
          \and
          P. Kervella\inst{2, 3}
          \and
          A. Gallenne\inst{4}
        \and
        L. Szabados\inst{5}
        \and
        R. I. Anderson\inst{6}\thanks{Swiss National Science Foundation Fellow}
        \and
        J.-B.~Le~Bouquin\inst{7}
         }

   \institute{European Southern Observatory, Alonso de C\'{o}rdova 3107, Casilla 19001, Santiago 19, Chile\\
         \email{joanne.breitfelder@obspm.fr}
        \and
        LESIA (UMR 8109), Observatoire de Paris, PSL, CNRS, UPMC, Univ. Paris-Diderot, 5 pl. Jules Janssen, 92195 Meudon, France
         \and
        Unidad Mixta Internacional Franco-Chilena de Astronom\'{i}a, CNRS/INSU, France (UMI 3386) and
        Departamento de Astronom\'{i}a, Universidad de Chile, Camino El Observatorio 1515, Las Condes, Santiago, Chile 
        \and
         Universidad de Concepci\'{o}n, Departamento de Astronom\'{i}a, Casilla 160-C, Concepci\'{o}n, Chile
        \and 
        Konkoly Observatory of the Hungarian Academy of Sciences, H-1121 Budapest, Konkoly Thege Str. 15-17, Hungary
        \and 
        Department of Physics and Astronomy, The Johns Hopkins University, 3400 N. Charles St, Baltimore, MD 21218, USA
        \and
        UJF-Grenoble 1 / CNRS-INSU, Institut de Plan\'{e}tologie et d'Astrophysique de Grenoble (IPAG) UMR 5274, Grenoble, France
          }

        \date{Received date / Accepted date}

        \abstract
        {The distance to pulsating stars is classically estimated using the parallax-of-pulsation (PoP) method, which combines spectroscopic radial velocity (RV) measurements and angular diameter (AD) estimates to derive the distance of the star. A particularly important application of this method is the determination of Cepheid distances in view of the calibration of their distance scale. However, the conversion of radial to pulsational velocities in the PoP method relies on a poorly calibrated parameter, the projection factor ($p$-factor).}
        {We aim to measure empirically the value of the $p$-factors of a homogeneous sample of nine bright Galactic Cepheids for which trigonometric parallaxes were measured with the Hubble Space Telescope (HST) Fine Guidance Sensor by Benedict et al. (2007).}
        {We use the {\tt SPIPS} algorithm, a robust implementation of the PoP method that combines photometry, interferometry, and radial velocity measurements in a global modeling of the pulsation of the star. We obtained new interferometric angular diameter measurements using the PIONIER instrument at the Very Large Telescope Interferometer (VLTI), completed by data from the literature. Using the known distance as an input, we derive the value of the $p$-factor of the nine stars of our sample and study its dependence with the pulsation period.} 
        {We find the following $p$-factors: $p = 1.20 \pm 0.12$ for RT~Aur, $p = 1.48 \pm 0.18$ for T~Vul, $p = 1.14 \pm 0.10$ for FF~Aql, $p = 1.31 \pm 0.19$ for Y~Sgr, $p = 1.39 \pm 0.09$ for X~Sgr, $p = 1.35 \pm 0.13$ for W~Sgr, $p = 1.36 \pm 0.08$ for $\beta$~Dor, $p = 1.41 \pm 0.10$ for $\zeta$~Gem, and $p = 1.23 \pm 0.12$ for $\ell$~Car.}
        {The values of the $p$-factors that we obtain are consistently close to $p=1.324 \pm 0.024$. We observe some dispersion around this average value, but the observed distribution is statistically consistent with a constant value of the $p$-factor as a function of the pulsation period ($\chi^2 = 0.669$). The error budget of our determination of the $p$-factor values is presently dominated by the uncertainty on the parallax, a limitation that will soon be waived by Gaia.}

        \keywords{Stars: variables: Cepheids, Techniques: interferometric, Methods: observational, Stars: distances}

\maketitle

\section{Introduction}

Cepheids are remarkable among variable stars for the tight relationship between their pulsation period and intrinsic luminosity (the Leavitt law; \citeads{1912HarCi.173....1L}). This empirical relation makes Cepheids very useful as primary distance indicator. Indeed, their brightness and number make them easily observable in the Milky Way, the Magellanic Clouds, and up to approximately 100\,Mpc. They are therefore a key element of the extragalactic cosmic ladder, and an accurate calibration of this law is fundamental.

A common way to estimate Cepheid distances is the parallax-of-pulsation (PoP) method, which relies on the comparison of the linear amplitude of the pulsation (derived from spectroscopic radial velocities) and its angular amplitude (from interferometry, or surface brightness-color relations). The PoP technique requires the translation of the spectroscopic radial velocity (hereafter RV) integrated over the disk of the star into a pulsation velocity (the velocity of the stellar photosphere). This conversion is achieved through a parameter, the projection factor ($p$-factor), whose calibration is currently uncertain at a 5 to 10\% level. Unfortunately, there is a full degeneracy between the $p$-factor and the derived distance, and this results in a global, systematic uncertainty of 5 to 10\% on the Cepheid distance scale calibrated using the PoP technique.
\citetads{2015arXiv151001940M} recently developed a new version of the PoP technique: the {\tt SPIPS} algorithm. This implementation is particularly robust as it is based on the full set of available observational constraints: multicolor photometry, RVs, and interferometric AD measurements. Previous PoP implementations rely only on two-color photometry and RVs, and they are therefore more prone to biases due to peculiar atmospheric effects (e.g., around the rebound phase), reddening, or circumstellar envelopes (CSEs). The {\tt SPIPS} algorithm relies on three assumptions:
\begin{enumerate}
\item Cepheids are pulsating on a radial mode, which is known to be true for most of them.
\item The angular size estimates (through interferometry and/or photometry) and the linear size measurement (from the integration of the RV curve) correspond to the same layer in the star. This is, in practice, not exactly the case, as the line-forming region is naturally located above the photosphere. In the present study, we consider velocities derived from a cross-correlation of the spectra, which represent an average altitude in the atmosphere, but do not match the photosphere exactly. The present calibration of the $p$-factor with {\tt SPIPS} implicitly includes this effect.
\item Cycle-to-cycle modulation in the amplitude of pulsation is sufficiently small: this is what allows us to combine data from different epochs. It has been reported recently that this is not entirely true for some Cepheids (\citeads{2014A&A...566L..10A}, \citeads{2015MNRAS.446.4008E}). However, these effects are only a second order contribution in error budget, and concern only one of the Cepheids studied, $\ell$~Car.
\end{enumerate}
A  calibration of the Leavitt law at a 1\% level requires unbiased distance measurements to calibrate Cepheids at the same level. It has been shown that this is a reachable goal with {\tt SPIPS}, under the condition that we calibrate the $p$-factor with sufficient accuracy.

Cepheids with a distance already known (e.g., {\it HST} parallax, light echoes, orbital parallax, etc.) allow us to break the degeneracy between the distance and the $p$-factor and study the possible correlation of this factor with the pulsation period (or other stellar parameters). We therefore take advantage of the parallaxes measured by \citetads{2007AJ....133.1810B} for nearby Cepheids to apply the {\tt SPIPS} algorithm and derive their $p$-factor values.
We present in Sect.~\ref{Observations_and_data_processing} our new VLTI/PIONIER interferometric AD measurements and the complementary datasets collected from the literature. Section~\ref{SPIPS} is dedicated to a brief description of the {\tt SPIPS} algorithm. We review our main results star-by-star in Sect.~\ref{Results}, and then discuss the resulting $p$-factor values in Sect.~\ref{Discussion}.

\section{Observations and data processing}
\label{Observations_and_data_processing}

\subsection{VLTI/PIONIER long-baseline interferometry}

In the past few years, we have led a large program of interferometric observations with the four-telescope beam-combiner PIONIER, installed at the VLTI at Cerro Paranal Observatory (Chile).
We present new data obtained for five classical Cepheids: X~Sgr, W~Sgr, $\zeta$~Gem, $\beta$~Dor, and $\ell$ Car. The observations were undertaken using the four 1.8~meter relocatable Auxiliary Telescopes in the largest available configuration. The largest baselines allow us to reach higher spatial frequencies, which are needed in this program since the AD of our targets is typically between 1 and 3~milliarcseconds (mas). The journal of the observations is summarized in Table~\ref{journal}.
For almost all the observations carried out in 2014 we used the SMALL dispersion mode of PIONIER allowing us to observe in three spectral channels of the $H$~band (1.59~$\mu$m, 1.67~$\mu$m, and 1.76~$\mu$m) and corresponding to a low spectral resolution of $R \sim 40$. We used the LARGE mode for the observations of the bright star $\ell$~Car, in which the light is dispersed over seven spectral channels of the $H$~band (1.52~$\mu$m, 1.55~$\mu$m, 1.60~$\mu$m, 1.66~$\mu$m, 1.71~$\mu$m, 1.76~$\mu$m, and 1.80~$\mu$m).
PIONIER was upgraded with a new detector between ESO periods 93 and 94. Consequently, all observations of 2015 were carried out using the new GRISM dispersion mode covering six spectral channels of the $H$~band (1.53~$\mu$m, 1.58~$\mu$m, 1.63~$\mu$m, 1.68~$\mu$m, 1.73~$\mu$m, and 1.78~$\mu$m), giving an equivalent spectral resolution of $R \sim 45$.
We alternated observations of each Cepheid  with observations of two different calibrator stars that were generally smaller in diameter to reach higher visibilities and not more than 5$^\circ$ away from the science target. The main characteristics of the calibrators are given in Table~\ref{calib}. They were all selected from \citetads{2005A&A...433.1155M} and the JMMC tool {\tt SearchCal} (\citeads{2010yCat.2300....0L}, \citeads{2006A&A...456..789B}, \citeads{2011A&A...535A..53B}).

\begin{figure}
        \includegraphics[width=\hsize]{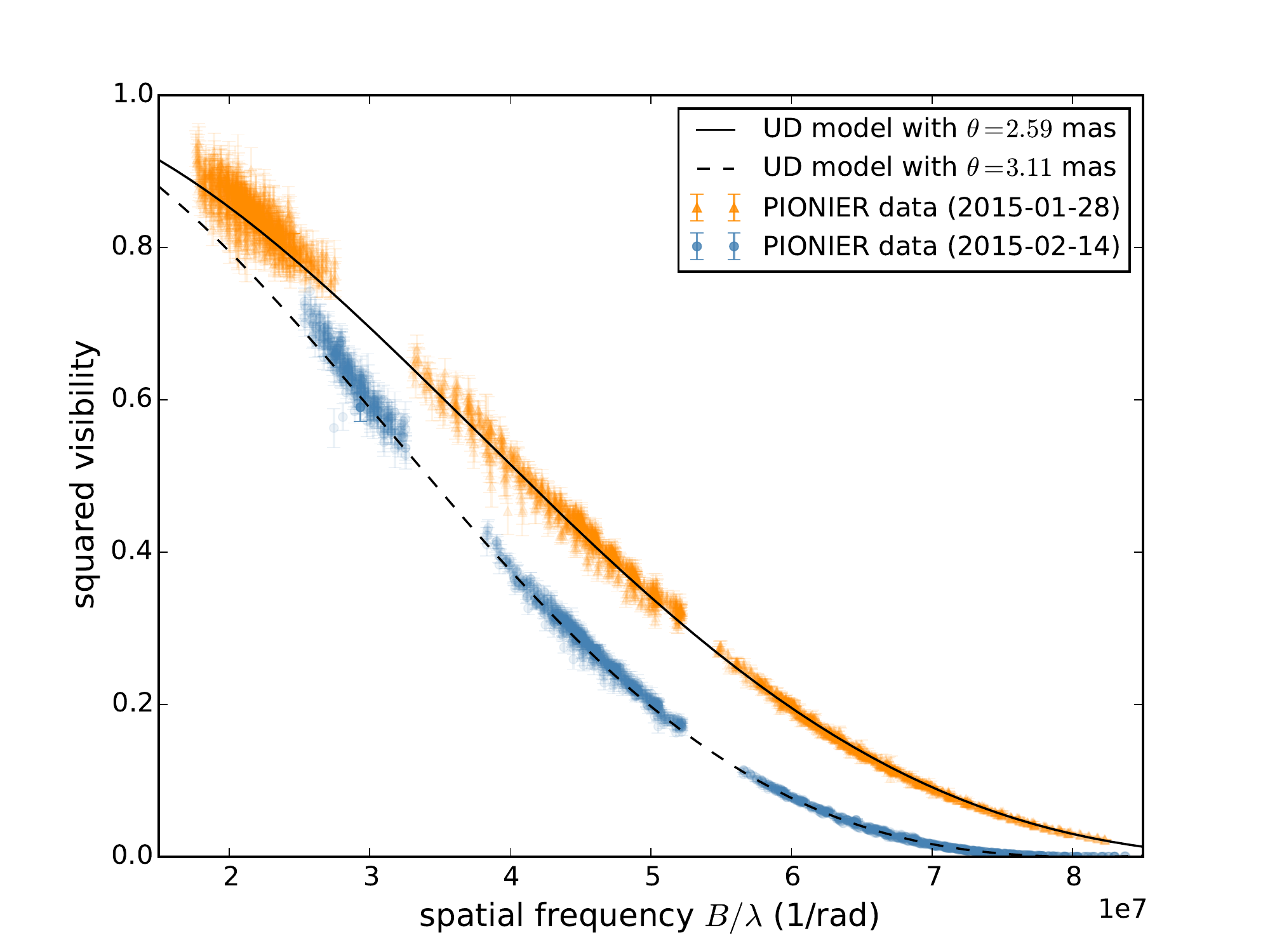}
        \caption{Squared visibilities measured with PIONIER at the minimum and  maximum diameters for $\ell$~Car. The data are fitted with a uniform disk model leading to the diameters of $\theta_{\rm{min}} = 2.59$~mas (MJD=57051, in orange) and $\theta_{\rm{min}} = 3.11$~mas (MJD=57068, in blue).}
        \label{lcar_vis}
\end{figure}

The raw data were reduced through the {\tt pndrs} data reduction software of PIONIER \citepads{2011A&A...535A..67L}, which provided us with calibrated squared visibilities and closure phases. We then adjusted these data with a uniform disk (UD) model using the \texttt{LITPro}\footnote{available at \url{http://www.jmmc.fr/litpro}} software \citepads{2008SPIE.7013E..44T} to retrieve UD angular diameters.
For each Cepheid, we obtain between 3 and 6 epochs that complement the literature data and provide a satisfactory coverage of the pulsation cycle. Of the stars observed, only $\zeta$~Gem does not have its diameter curve fully covered by our observations. 
The resulting ADs are listed in Table~\ref{PIONIER_data}. For each Cepheid, we specify the Modified Julian Date (MJD) of the observations, from which the pulsation phase is derived using the ephemeris in Table~\ref{ephemeris}, the UD angular diameter, the statistical error bar given by the model fitting, and the systematic error which is defined as the mean error on the calibrators diameters.
We also indicate the $\chi^2$ of the UD model fitting to show the consistency of the statistical error, which is small thanks to the high number of single visibility measurements. The Cepheid $\ell$~Car has been observed in the framework of two different programs. The three first epochs are of a less good quality and the UD model fit consequently leads to higher $\chi^2$, while the other data points result from longer observations (about five hours of observation per night) and have therefore very small uncertainties (see Anderson et al. in prep.). An example of the very good quality visibility curves obtained for the minimum and maximum diameter of $\ell$~Car are shown in Fig.~\ref{lcar_vis}.
For the SPIPS model fit (see \citetads{2015arXiv151001940M}; see also Sect.~\ref{SPIPS}), the uniform disk ADs are converted to limb darkening (LD) values using SATLAS spherical atmosphere models \citepads{2013A&A...554A..98N}.

We used the {\tt CANDID}\footnote{\url{https://github.com/amerand/CANDID}} tool \citepads{2015A&A...579A..68G} to check all our interferometric data (considering both visibilities and phase closures) for the presence of close companions (located within $\approx 50$\,mas of the Cepheid). At the detection level of {\tt CANDID} (about 1\% in flux ratio), we did not find any significant signal at more than $3\sigma$, we therefore conclude that our diameter measurements are not biased by the contribution of resolved companions. A precise study of binarity with {\tt CANDID} would actually require more time of integration on each Cepheid. Data dedicated to diameter measurement are in general less numerous.
The raw data are all available on the ESO Archive and the reduced data are available from the Jean-Marie Mariotti Center \texttt{OiDB} service\footnote{\url{http://oidb.jmmc.fr/index.html}}. They result from a basic calibration and may be slightly different from the data presented here, since we made our own calibration, excluding in particular the observations undertaken under bad conditions or degraded by instrumental issues.
We completed our sample of interferometric AD measurements with values from \citetads{2004A&A...416..941K}, \citetads{2002ApJ...573..330L}, \citetads{2009MNRAS.394.1620D}, \citetads{2012A&A...541A..87G} and \citetads{trove.nla.gov.au/work/31422632}.

\begin{table}
    \caption{Journal of our new PIONIER observations}
    \label{journal}
    \centering
    \renewcommand{\arraystretch}{1.2}
    \begin{tabular}{*{4}{c}}
        \hline
        date & ATs config. & Disp. & Seeing \\
        \hline
        \hline
        2014-04-02 & A1-G1-K0-J3 & LARGE & 0.7-2.2 \\
        2014-04-04 & A1-G1-K0-J3 & SMALL & 0.7-1.7 \\
        2014-05-07 & A1-G1-K0-J3 & SMALL & 0.4-1.9 \\
        2014-07-28 & A1-G1-K0-J3 & SMALL & 0.4-1.6 \\
        2014-07-30 & A1-G1-K0-J3 & SMALL & 0.4-1.6 \\
        2014-08-01 & A1-G1-K0-J3 & SMALL & 0.7-3.5 \\
        2014-08-04 & A1-G1-K0-J3 & SMALL & 0.7-3.0 \\
        2014-08-19 & A1-G1-K0-J3 & SMALL & 0.6-3.0 \\
        2014-08-24 & A1-G1-K0-J3 & SMALL & 0.5-1.2\\
        2015-01-13 & A1-G1-K0-I1 & GRISM & 0.6-1.5 \\
        2015-01-14 & A1-G1-K0-I1 & GRISM & 0.5-1.6 \\
        2015-01-15 & A1-G1-K0-I1 & GRISM & 0.5-1.7 \\
        2015-01-16 & A1-G1-K0-I1 & GRISM & 0.6-2.0 \\
        2015-01-26 & A1-G1-K0-I1 & GRISM & 0.4-1.9 \\
        2015-01-27 & A1-G1-K0-I1 & GRISM & 0.6-2.7 \\
        2015-01-28 & A1-G1-K0-I1 & GRISM & 0.4-1.4 \\
        2015-01-29 & A1-G1-K0-I1 & GRISM & 0.8-1.6 \\
        2015-01-30 & A1-G1-K0-I1 & GRISM & 0.5-1.9 \\
        2015-02-04 & D0-H0-G1-I1 & GRISM & Unknown \\
        2015-02-05 & D0-H0-G1-I1 & GRISM & Unknown \\
        2015-02-12 & A1-G1-K0-J3 & GRISM & Unknown \\
        2015-02-13 & A1-K0-J3 & GRISM & 0.7-2.8 \\
        2015-02-14 & A1-G1-K0-J3 & GRISM & Unknown \\
        2015-02-15 & A1-G1-K0-J3 & GRISM & 0.5-1.8 \\
        2015-02-16 & A1-G1-K0-J3 & GRISM & Unknown \\
        2015-02-17 & A1-G1-K0-J3 & GRISM & 0.7-2.4 \\
        2015-02-18 & A1-G1-K0-J3 & GRISM & 0.9-3.0 \\
        2015-02-20 & A1-G1-K0-J3 & GRISM & 0.8-2.9 \\
        \hline
    \end{tabular}
    \tablefoot{PIONIER spectral dispersion setup: LARGE; dispersion over seven spectral channels of the $H$~band (1.52, 1.55, 1.60, 1.66, 1.71, 1.76, and 1.80~$\mu$m). SMALL; dispersion over three spectral channels of the $H$~band (1.59, 1.67, and 1.76~$\mu$m). GRISM; dispersion over six spectral channels of the $H$~band (1.53, 1.58, 1.63, 1.68, 1.73, and 1.78~$\mu$m).}
\end{table}

\begin{table*}
    \caption{PIONIER observations. We give here the mean MJD (defined as JD$-$2\,400\,000.5) of each observing night, the corresponding phase (taking $\phi_{0}$ at the maximum of luminosity in $V$), the range of baselines, the best uniform disk diameter adjusted on the squared visibility measurements, its uncertainty, and the reduced $\chi^2$ of the fit.}
    \label{PIONIER_data}
    \centering
    \renewcommand{\arraystretch}{1.2}
    \begin{tabular}{*{5}{c}}
        \hline
        MJD & Baselines (m) & Phase & $\theta_{UD} \pm \sigma_{\rm{stat}} \pm \sigma_{\rm{syst}}$ (mas) & $\chi^2$ \\
        \hline
        \hline
        \multicolumn{4}{c}{\emph{ X Sgr }} \\
            56867.1442 & 56.76 - 139.97 & 0.49 & 1.3428 $\pm$ 0.0025 $\pm$ 0.0595 & 2.85 \\
            56869.1692 & 56.76 - 139.97 & 0.78 & 1.2791 $\pm$ 0.0031 $\pm$ 0.0170 & 1.18 \\
            56871.1730 & 56.76 - 139.97 & 0.07 & 1.3185 $\pm$ 0.0034 $\pm$ 0.0170 & 1.07 \\
            56874.1975 & 56.76 - 139.97 & 0.50 & 1.3507 $\pm$ 0.0125 $\pm$ 0.0170 & 2.56 \\
            56894.1539 & 56.76 - 139.97 & 0.34 & 1.4098 $\pm$ 0.0033 $\pm$ 0.0170 & 1.11 \\
        \hline
        \multicolumn{4}{c}{\emph{ W Sgr }} \\
            56867.0798 & 56.76 - 139.97 & 0.53 & 1.1978 $\pm$ 0.0018 $\pm$ 0.0170 & 1.34 \\
            56869.0861 & 56.76 - 139.97 & 0.79 & 1.0778 $\pm$ 0.0036 $\pm$ 0.0170 & 2.80 \\
            56871.1161 & 56.76 - 139.97 & 0.06 & 1.0995 $\pm$ 0.0036 $\pm$ 0.0170 & 0.91 \\
            56874.1477 & 56.76 - 139.97 & 0.46 & 1.1606 $\pm$ 0.0044 $\pm$ 0.0170 & 2.45 \\
            56894.1044 & 56.76 - 139.97 & 0.09 & 1.1184 $\pm$ 0.0021 $\pm$ 0.0170 & 0.82 \\
            56889.1339 & 56.76 - 139.97 & 0.43 & 1.1711 $\pm$ 0.0047 $\pm$ 0.0170 & 1.02 \\
        \hline
        \multicolumn{4}{c}{\emph{ $\zeta$ Gem }} \\
            57038.2455 & 46.64 - 129.08 & 0.73 & 1.5372 $\pm$ 0.0025 $\pm$ 0.0425 & 1.97 \\
            57039.2786 & 46.64 - 129.08 & 0.83 & 1.5871 $\pm$ 0.0029 $\pm$ 0.0425 & 2.01 \\
            57071.1621 & 56.76 - 139.97 & 0.97 & 1.6047 $\pm$ 0.0020 $\pm$ 0.0690 & 1.17 \\
            57067.1395 & 56.76 - 139.97 & 0.57 & 1.5660 $\pm$ 0.0020 $\pm$ 0.0425 & 0.99 \\
        \hline
        \multicolumn{4}{c}{\emph{ $\beta$ Dor }} \\
            57036.0889 & 46.64 - 129.08 & 0.91 & 1.6857 $\pm$ 0.0016 $\pm$ 0.0120 & 2.41 \\
            57037.0627 & 46.64 - 129.08 & 0.01 & 1.7584 $\pm$ 0.0012 $\pm$ 0.0120 & 1.21 \\
            57038.0605 & 46.64 - 129.08 & 0.11 & 1.8160 $\pm$ 0.0010 $\pm$ 0.0120 & 2.98 \\
            57071.0192 & 56.76 - 139.97 & 0.46 & 1.7939 $\pm$ 0.0020 $\pm$ 0.0120 & 1.03 \\
            57074.0888 & 56.76 - 139.97 & 0.77 & 1.6022 $\pm$ 0.0040 $\pm$ 0.0120 & 1.78 \\
            57067.0280 & 56.76 - 139.97 & 0.05 & 1.7098 $\pm$ 0.0020 $\pm$ 0.0120 & 2.43 \\
        \hline
        \multicolumn{4}{c}{\emph{ $\ell$ Car }} \\
            56750.0604 & 56.76 - 139.97 & 0.41 & 3.1059 $\pm$ 0.0032 $\pm$ 0.0990 & 5.88 \\
            56752.1463 & 56.76 - 139.97 & 0.47 & 3.0803 $\pm$ 0.0028 $\pm$ 0.0575 & 6.76 \\
            56785.0895 & 56.76 - 139.97 & 0.40 & 3.1143 $\pm$ 0.0017 $\pm$ 0.0575 & 3.97 \\
            57049.2983 & 46.64 - 129.08 & 0.83 & 2.6383 $\pm$ 0.0003 $\pm$ 0.0160 & 0.03 \\
            57050.2604 & 46.64 - 129.08 & 0.85 & 2.6009 $\pm$ 0.0009 $\pm$ 0.0160 & 0.04 \\
            57051.3043 & 46.64 - 129.08 & 0.88 & 2.5999 $\pm$ 0.0004 $\pm$ 0.0160 & 0.03 \\
            57052.3017 & 46.64 - 129.08 & 0.91 & 2.6038 $\pm$ 0.0003 $\pm$ 0.0160 & 0.02 \\
            57053.3081 & 46.64 - 129.08 & 0.94 & 2.6156 $\pm$ 0.0004 $\pm$ 0.0160 & 0.02 \\
            57058.3484 & 41.03 - 82.48 & 0.08 & 2.8430 $\pm$ 0.0025 $\pm$ 0.0160 & 0.05 \\
            57059.3568 & 41.03 - 82.48 & 0.11 & 2.8913 $\pm$ 0.0035 $\pm$ 0.0160 & 0.05 \\
            57066.1018 & 56.76 - 139.97 & 0.30 & 3.0929 $\pm$ 0.0002 $\pm$ 0.0160 & 0.02 \\
            57068.2169 & 56.76 - 139.97 & 0.36 & 3.1117 $\pm$ 0.0003 $\pm$ 0.0160 & 0.03 \\
            57069.1280 & 56.76 - 139.97 & 0.39 & 3.1089 $\pm$ 0.0002 $\pm$ 0.0160 & 0.03 \\
            57070.1479 & 56.76 - 139.97 & 0.41 & 3.1104 $\pm$ 0.0003 $\pm$ 0.0160 & 0.03 \\
            57072.1099 & 56.76 - 139.97 & 0.47 & 3.0888 $\pm$ 0.0003 $\pm$ 0.0160 & 0.03 \\
        \hline
    \end{tabular}
\end{table*}

\begin{table}
    \caption{Properties of the interferometric calibrators used during our PIONIER observations. We indicate the uniform disk diameter in the $H$ band and the corresponding uncertainty.}
    \label{calib}
    \centering
    \renewcommand{\arraystretch}{1.2}
    \begin{tabular}{*{5}{c}}
        \hline
        Star & $m_{v}$ & $m_{H}$ & $\theta_{UD} \pm \sigma$ (mas) & \emph{Ref.} \\
        \hline
        \hline
         HD35199 & 7.2 & 4.11 & 0.854 $\pm$ 0.012 & (a) \\
         HD39608 & 7.35 & 3.96 & 0.939 $\pm$ 0.012 & (a) \\
         HD50607 & 6.55 & 4.59 & 0.594 $\pm$ 0.042 & (b) \\
         HD50692 & 5.76 & 4.51 & 0.604 $\pm$ 0.043 & (b) \\
         HD54131 & 5.49 & 3.22 & 1.356 $\pm$ 0.096 & (b) \\
         HD81101 & 4.8 & 2.66 & 1.394 $\pm$ 0.099 & (b) \\
         HD81502 & 6.29 & 3.24 & 1.23 $\pm$ 0.016 & (a) \\
         HD89805 & 6.3 & 2.87 & 1.449 $\pm$ 0.019 & (a) \\
         HD156992 & 6.36 & 3.12 & 1.24 $\pm$ 0.017 & (a) \\
         HD166295 & 6.68 & 2.93 & 1.266 $\pm$ 0.017 & (a) \\
         HD166464 & 4.98 & 2.68 & 1.434 $\pm$ 0.102 & (b) \\
         HD170499 & 7.73 & 3.25 & 1.235 $\pm$ 0.017 & (a) \\
        \hline
    \end{tabular}
    \tablefoot{References:
    (a) \citeads{2005A&A...433.1155M};
    (b) JMMC catalog of calibration sources}
\end{table}

\subsection{Radial velocity measurements from the literature}

The present study makes use of the following references providing RV data: \citetads{2014A&A...566L..10A}, \citetads{2005ApJS..156..227B}, \citetads{1994A&AS..108...25B}, \citetads{2002ApJS..140..465B}, \citetads{1990AJ.....99.1598E}, \citetads{1998AstL...24..815G}, \citetads{1998MNRAS.297..825K}, \citetads{2009A&A...502..951N}, \citetads{2005MNRAS.362.1167P}, and \citetads{2011A&A...534A..94S}.

The data coming from these different sources are consistent with each other, since almost all the RVs were determined with the same method (i.e., a cross-correlation of the spectra with a binary mask and a Gaussian fit of the resulting cross-correlation profile), and are given in the International Astronomical Union standard RV system.
Only the velocities of \citetads{2005MNRAS.362.1167P} have been treated differently and result from a measurement of the line bisector. A change in the measurement method has an effect on the amplitude of the RV curve, especially if the spectral lines become highly asymmetric during the pulsation \citepads{2006A&A...453..309N}. Nevertheless, we needed these data to get a sufficiently complete coverage of the RV curve of $\beta$~Dor and W~Sgr. Fortunately, none of these stars shows a significant amplitude modulation between these data and the other data sets (based on cross-correlation) that were used jointly. 
The CORAVEL data from \citetads{1994A&AS..108...25B} and \citetads{2002ApJS..140..465B} are given with an offset of +0.4~km\,s$^{-1}$ compared to IAU standard, while the zero point of \citetads{1998AstL...24..815G} data is given between +0.5 and +1.5~km\,s$^{-1}$ because different instruments were used in the observing campaign. Although we did not use it in the present study, we underline that all the CORAVEL zero points have been recently listed by \citetads{2015AJ....150...13E}.
Except for $\zeta$~Gem, $\beta$~Dor, T~Vul, and Y~Sgr, we cannot see vertical shifts of the RV curves coming from different data sets. For the stars mentioned above, we simply corrected the different RV curves so that, for each
author, the mean value of the model would coincide with $V_{\rm{mean}} = 0$ km\,s$^{-1}$. This process allows us to "clean" the curve for possible biases like zero point uncertainties or Keplerian motion due to a companion and to keep only the pulsation component. A discussion about the offsets between different RV data sets is presented in \citetads{2000MNRAS.314..420K}. \citetads{1998MNRAS.297..825K} gives two different values for the RVs, resulting from the cross-correlation of two different parts of the spectra. We have opted to keep the mean value of both measurements. 
Finally, we consider a systematic error of $\pm 0.3$\,km\,s$^{-1}$
that we quadratically add to the uncertainties of all the RV data to take all the systematic effects due to the combination of different data sets into account.
The Cepheids of our sample have a reasonably good coverage in RV, which is essential for a proper estimation of the radius curve (Sect.~\ref{SPIPS}).

\subsection{Photometry from the literature}

The present study makes use of an extensive collection of optical and near-infrared (hereafter IR) light curves. We use $BV$ photometric data from \citetads{1998MNRAS.297..825K}, \citetads{1997PASP..109..645B}, \citetads{2008yCat.2285....0B}, \citetads{1977MmRAS..83...69D}, \citetads{1975ApJS...29..219M}, \citetads{1984ApJS...55..389M}, \citetads{1992MNRAS.255..486S}, \citetads{1977CoKon..70....1S}, \citetads{1981CoKon..77....1S}, and \citetads{1991CoKon..96..123S}.
Most of these magnitudes are expressed in the standard Johnson-Morgan-Cousins system, therefore, we fit them with the Johnson and Cousins filters provided by the General Catalog of Photometric Data (GCPD) and revised by \citetads{2015PASP..127..102M}\footnote{\url{http://svo2.cab.inta-csic.es/theory/fps3/index.php}}. We also use data from the {\it Hipparcos} and {\it Tycho} catalogs \citepads{1997ESASP1200.....E}, which we fit with the dedicated {\it Hipparcos} and {\it Tycho} $B$ and $V$ band filters, also revised by \citetads{2015PASP..127..102M}. Finally, we use Geneva magnitudes from \citetads{1994A&AS..108....9B} and \citetads{2002ApJS..140..465B} (only in the $V$ band), which were fitted with the suited Geneva $V$ band filter provided by the {\it Spanish Virtual Observatory}.
We do not use the photometry in the $R$ and $I$ bands provided by some of these authors, since the detector's quantum efficiency is generally uncertain in this wavelength range and the filter+detector effective bandpass is therefore poorly defined. As a consequence, these data tend to degrade the quality of the overall fit. In any event, the temperature and  reddening information are mainly contained in the $B$ and $V$ bands, while the envelope is seen in IR. 
We also include photometry in the IR $JHK$ bands, which is less sensitive to the interstellar reddening and more sensitive to the effective temperature. We gathered data from \citetads{1997PASP..109..645B}, \citetads{2011ApJS..193...12M}, and \citetads{1984ApJS...54..547W}, which are all given in the CTIO photometric system; and from \citetads{1980SAAOC...1..163L}, \citetads{2008MNRAS.386.2115F}, and \citetads{1992A&AS...93...93L}, which we converted from the SAAO to the CTIO systems through the laws given in \citetads{1990MNRAS.242....1C}\footnote{The relations between SAAO and other systems given by Carter et al. 1990 are summarized on the Asiago Database on Photometric Systems webpage (\url{http://ulisse.pd.astro.it/Astro/ADPS/Systems/Sys_137/fig_137.gif})}.

Most authors give a standard deviation of 0.01 to 0.02 magnitudes for the individual measurements. The data from {\it Hipparcos} have very small error bars. To give them an equivalent weight in the fitting process, we multiplied all of the uncertainties by an arbitrary factor of $3$, which allows us to obtain a reduced $\chi^2$ close to 1 for the fit of this particular data set.
To take the different instrumental calibrations into account, we added a systematic uncertainty of 0.02 magnitudes to all our photometric data. This value is consistent with the average offset generally observed when combining data from different instruments and magnitude systems (see, for instance, \citeads{1997PASP..109..645B}).
The references of the data used for our nine Cepheids are summarized in Table~\ref{data}. All Cepheids have an excellent phase coverage in all the selected optical and IR bands.

%



%

\begin{table}
    \caption{Data used on each star to apply the {\tt SPIPS} method. We gathered  all the best quality radial velocities, photometry in bands $BVRI$ and $JHK$, and interferometric diameters (Diam.) from the literature.}
    \label{data}
    \centering
    \renewcommand{\arraystretch}{1.3}
    \begin{tabular}{*{5}{c}}
        \hline
        Star                    & RVs           & $BV$                  & $JHK$           & Diam. \\
         \hline
         \hline
        $\beta$ Dor     & d, h, i               & d, l, n, q, u         & v, x            & A, C, F \\
        $\zeta$ Gem     & c, f, g, h, j & l, g, p, q, s, u, m   & w                      & B, C \\
        X Sgr           & j                     & d, l, p, q, u         & w, z            & A, C \\
        Y Sgr           & d, h, j               & d, l, p, q, u         & z                       & - \\
        W Sgr           & c, i                  & l, p, q, u, m         & z                       & A, C \\
        FF Aql          & e, f, g               & l, p, r, t, u         & z                       & E \\
        RT Aur          & f, g                  & k, l, g, p, r, u              & j, y                    & - \\
        T Vul           & b, c, g               & k, l, g, p, r, u, m   & j, z                    & E \\
        $\ell$ Car              & a, d, h, i            & d, l, o, u, n         & x                       & A, C, D \\
        \hline
    \end{tabular}
    \tablefoot{{\it References:} (a)~\citetads{2014A&A...566L..10A}; (b)~\citetads{2005ApJS..156..227B}; (c)~\citetads{1994A&AS..108...25B}; (d)~\citetads{2002ApJS..140..465B}; (e)~\citetads{1990AJ.....99.1598E}; (f)~\citetads{1998AstL...24..815G}; (g)~\citetads{1998MNRAS.297..825K};
(h)~\citetads{2009A&A...502..951N}; (i)~\citetads{2005MNRAS.362.1167P}; (j)~\citetads{2011A&A...534A..94S}; (k)~\citetads{1997PASP..109..645B}; (l)~\citetads{2008yCat.2285....0B}; (m)~\citetads{1994A&AS..108....9B}; (n)~\citetads{1977MmRAS..83...69D}; (o)~\citetads{1975ApJS...29..219M}; (p)~\citetads{1984ApJS...55..389M};
(q)~\citetads{1992MNRAS.255..486S}; (r)~\citetads{1977CoKon..70....1S}; (s)~\citetads{1981CoKon..77....1S};
(t)~\citetads{1991CoKon..96..123S}; (u)~\citetads{1997ESASP1200.....E}; (v)~\citetads{1980SAAOC...1..163L};
(w)~\citetads{2008MNRAS.386.2115F}; (x)~\citetads{1992A&AS...93...93L}; (y)~\citetads{2011ApJS..193...12M};
(z)~\citetads{1984ApJS...54..547W}; (A)~VINCI/VLTI and FLUOR/IOTA data \citepads{2004A&A...416..941K};
(B)~PTI data \citepads{2002ApJ...573..330L}; (C)~PIONIER data ({\it present work}); 
(D)~SUSI data \citepads{2009MNRAS.394.1620D}; (E)~FLUOR/CHARA data \citepads{2012A&A...541A..87G};
(F)~SUSI data \citepads{trove.nla.gov.au/work/31422632}.}
\end{table}

\section{The {\tt SPIPS} algorithm \label{SPIPS}}

To reproduce our complete observational data set, we use the {\it Spectro-Photo-Interferometry of Pulsating Stars} modeling tool ({\tt SPIPS}; \citeads{2015arXiv151001940M}), inspired from the classical PoP technique (commonly known as "Baade-Wesselink").
The general idea of this method is to compare the linear and angular variations of the Cepheid diameters to retrieve the distance. The {\tt SPIPS} code can take  all the different types of data and observables that can be found in the literature into account, in particular, magnitudes and colors in all optical and IR bands and filters, RVs, and interferometric ADs.
The resulting redundancy in the observables ensures a higher level of robustness. For instance, the AD is constrained by both the interferometry and photometry (via the use of atmospheric models).
The {\tt SPIPS} code also allows us to fit an excess in $K$ and $H$ band, to bring out the possible presence of a CSE. It also outputs the color excess $E(B-V)$, derived through the reddening law from \citetads{1999PASP..111...63F}, considering the classical Galactic value of the total-to-selective absorption $R_{\rm{V}}=3.1$. 

It is necessary to set the value of the $p$-factor (also abbreviated as $p$ in the following) used to convert the spectral RVs into photospheric pulsation velocities through $V_{\rm{puls}}=p\,V_{\rm{rad}}$. The $p$-factor is fully degenerate with the distance in the PoP technique (including {\tt SPIPS}). In fact, $p$ and $d$ are symmetrical in the fitting process, and only the ratio $p/d$ can be derived unambiguously unless one of these two parameters can be determined independently and input in {\tt SPIPS} as a fixed parameter.
The $p$-factor is also sensitive to the spectral lines that are considered, since they are all formed in different layers of the atmosphere and do not pulsate at the exact same velocity. Observing in different lines (e.g., different line-forming regions) consequently leads to different $p$-factors.
In the present study, we mainly use cross-correlation velocities that allow us to average out the differential atmospheric effects. The method used to derive the RVs is an important point in the PoP method, as the curves obtained with different techniques can have more than 5\% difference in amplitude \citepads{2007A&A...471..661N}. It is important to stress that the results from the present study are suited for the cross-correlation method and a Gaussian fit of the cross-correlation profile. 

We selected nine Cepheids whose parallax has been measured by \citetads{2007AJ....133.1810B} with the Fine Guidance Sensor (FGS) on board the {\it Hubble Space Telescope}: \object{RT Aur}, \object{T Vul}, \object{FF Aql}, \object{Y Sgr}, \object{X Sgr}, \object{W Sgr}, \object{$\beta$ Dor}, \object{$\zeta$ Gem,} and \object{$\ell$ Car}.
Knowing the distance, we can break the degeneracy of the PoP method and deduce the value of their $p$-factor, as already carried out on the prototype classical Cepheid $\delta$~Cep by \citetads{2005A&A...438L...9M} and on the type II Cepheid $\kappa$~Pav by \citetads{2015A&A...576A..64B}.
These studies give values of  $1.27 \pm 0.06$ and $1.26 \pm 0.07$, respectively, for the $p$-factor.
For each Cepheid, we fit the RV curves using spline functions defined by semifixed nodes. Although this method is numerically less stable than Fourier series, it leads to smoother models and avoids the introduction of unphysical oscillations when the data are too dispersed or not dense enough.
The photometry curves are fitted with Fourier series. This does not introduce spurious oscillations thanks to the large quantity and good phase coverage of photometric data collected for each star.
To take into account  each main observable (RV, interferometry, IR photometry, and optical photometry) in a balanced way, we allocate to these different data sets the same weight in the fitting process. We do this by multiplying the error bars by a factor inversely proportional to the number of data points contained in that observable group.
For each Cepheid, we set a reference MJD taken as close as possible to the center of the time interval covered by the data and corresponding to a maximum of luminosity.
We fit both the period and a linear variation $dP/dt$ with the {\tt SPIPS} code; the best parameters are those allowing the smallest dispersion of the data.
This approach is different from the usual study of the O-C diagram and does not always lead to identical results (Sect.~\ref{Results}). The final ephemerides used to phase the data are given in Table~\ref{ephemeris}. We indicate the reference date, the period and its variation, and the corresponding crossing of the Cepheid in the instability strip (deduced from the predictions of \citeads{2014AstL...40..301F}). The table also gives the epoch range covered by the data, which is  relevant information for the calculation of the $dP/dt$ variation.
The results are described star-by-star in Sect.~\ref{Results} and the graphics resulting from the {\tt SPIPS} modeling are shown in the annexes. The values of all the best-fit parameters are given in Table~\ref{results}, where we indicate both the statistical and systematic errors. In the case of the $p$-factor, the systematic error is due to the parallax. For the temperature and reddening, the systematic error has been set by running a "jackknife" algorithm on the photometric data of $\delta$~Cep, a star that has been extensively studied in \citetads{2005A&A...438L...9M} and \citetads{2015arXiv151001940M}. This method leads to uncertainties of 0.016 for reddenings and 50~K for temperatures. When no interferometric diameters were available, we considered a systematic error of 2\% on the diameters \citepads{2004A&A...428..587K}.

\begin{table*}
    \caption{Ephemeris used to phase the data of each Cepheid. We also give the epoch range covered by the data for each Cepheid.}
    \label{ephemeris}
    \centering
    \renewcommand{\arraystretch}{1.2}
    \begin{tabular}{*{6}{c}}
        \hline
        Star & $\rm{MJD}_{\rm{0}}$ & Period (days) & dP/dt (sec/yr) & Crossing & Epoch range \\
        & & (days) & (sec/yr) & & (yrs) \\
        \hline
        \hline
         RT Aur & 48027.678 & 3.728305 $\pm$ 0.000005 & 0.124 $\pm$ 0.036 & 3 & 36 \\
         T Vul & 49134.074 & 4.435424 $\pm$ 0.000005 & 0.060 $\pm$ 0.035 & 3 & 25 \\
         FF Aql & 45912.675 & 4.470848 $\pm$ 0.000010 & $-$0.140 $\pm$ 0.036 & 2 & 66 \\
         Y Sgr & 47303.129 & 5.773383 $\pm$ 0.000009 & 0.016 $\pm$ 0.048 & 3 & 30 \\
         X Sgr & 49310.835 & 7.012770 $\pm$ 0.000012 & 0.371 $\pm$ 0.098 & 3 & 37 \\
         W Sgr & 48257.806 & 7.594984 $\pm$ 0.000009 & 0.331 $\pm$ 0.111 & 3 & 37 \\
         $\beta$ Dor & 49133.243 & 9.842675 $\pm$ 0.000019 & $-$0.084 $\pm$ 0.149 & 2 & 42 \\
         $\zeta$ Gem & 49134.561 & 10.149806 $\pm$ 0.000017 & $-$1.238 $\pm$ 0.144 & 2 & 43 \\
         $\ell$ Car & 47774.310 & 35.551609 $\pm$ 0.000265 & 27.283 $\pm$ 0.984 & 3 & 42 \\
        \hline
    \end{tabular}
\end{table*}

\section{Results}
\label{Results}

\subsection{RT Aur}


RT~Aur is a very short period Cepheid (3.7 days). Its cycle-to-cycle photometric variations were recently studied by \citetads{2015MNRAS.446.4008E}, who found a high repeatability in amplitude, but a slow drift of 0.000986 days per century (0.852 s/yr) in pulsation period. \citetads{2007PASP..119.1247T} propose a much lower value of $0.082 \pm 0.012$~s/yr, which is closer to our own value of $0.124 \pm 0.036$~s/yr. This period change is that expected for a Cepheid crossing the instability strip for the third time \citepads{2014AstL...40..301F}. \citetads{2007PASP..119.1247T} observed a sinusoidal trend in the O-C diagram, which is interpreted as a light-time effect produced by a long period orbit companion. \citetads{2015MNRAS.446.4008E} also reported a slight decrease in $v_{\gamma}$, but they did not reach a conclusion about the presence of a companion.
\citetads{2015A&A...579A..68G} detected the companion from CHARA/MIRC interferometric observations via the {\tt CANDID} code. The data revealed a very close companion lying only 2.1\,mas away from the Cepheid. This finding is, however, unconfirmed and demands further studies.
\citetads{2015A&A...579A..68G} published a UD diameter of $0.699 \pm 0.011$ at $\phi=0.32$, which is consistent with the value found in the present study at the same phase. \citetads{2008MNRAS.389.1336K} and \citetads{2007AJ....133.1810B} gave respective color excesses of $E(B-V) = 0.050 \pm 0.036$ and $0.051$, which are both consistent with the color excess resulting from our {\tt SPIPS} fit ($E(B-V) = 0.048 \pm 0.016$).
We find a $p$-factor of $1.20 \pm 0.08_{\rm{stat}} \pm 0.09_{\rm{sys}}$, which is in agreement at a 2$\sigma$ level with most values deduced from published period-$p$ relations. The value found in the present study and the value from \citetads{2009A&A...502..951N} agree within their error bars. The final {\tt SPIPS} adjustment is shown in Fig.~\ref{RTAur}.

\subsection{T Vul}


T~Vul is a bright, short-period (4.4~days) northern Cepheid. For this star, we corrected the different RV data sets from their mean value (calculated using the same model for each author): $-0.988 \pm 0.032$~km\,s$^{-1}$ for \citetads{2005ApJS..156..227B}, $-2.664 \pm 0.033$~km\,s$^{-1}$ for \citetads{1994A&AS..108...25B}, and $-0.759 \pm 0.030$~km\,s$^{-1}$ for \citetads{1998MNRAS.297..825K}. A simple linear period variation did not allow us to phase  the CHARA/FLUOR interferometric diameters properly from \citetads{2012A&A...541A..87G} with the rest of the data. We therefore kept the pulsation phases given by the authors and added an offset of $\phi = -0.2328$ to reach the best phase agreement. 
The interferometric data in \citetads{2012A&A...541A..87G} lead to a mean diameter of $0.629 \pm 0.013$~mas for T~Vul, consistent with our {\tt SPIPS} diameter of $0.607 \pm 0.012$~mas. Like us, \citetads{2012A&A...541A..87G} use the parallax from \citetads{2007AJ....133.1810B} ($\pi = 1.90 \pm 0.23$, $d = 526.31 \pm 63.71$~pc), and deduce a linear radius $R = 35.6 \pm 4.4~R_{\odot}$, coherent with our own value of $35.39 \pm 0.07_{\rm{stat}} \pm 4.98_{\rm{sys}} ~R_{\odot}$.
A faint spectroscopic companion of type A0.8 V has been discovered in the {\it International Ultraviolet Explorer} ({\it IUE}) observations from \citetads{1992AJ....104..216E}, and \citetads{2015A&A...579A..68G} did not detect any companion with a spectral type earlier than B9V within 50\,mas.
\citetads{2008MNRAS.389.1336K},  \citetads{2007AJ....133.1810B} and \citetads {1992AJ....104..216E} published similar values of  $0.068 \pm 0.015$, $0.064$ and $0.060$, respectively, for the color excess. Our study leads to a lower value of $0.019 \pm 0.016$. Various authors agree that the pulsation period of T~Vul is subject to a slight decrease with time. \citetads{2006OEJV...46....1M} mentioned that the change rate is in the interval of $-0.25 \pm 0.13$~s/yr with a probability of 99\%, while \citetads{1998JAVSO..26..101T} suggests a similar value of $-0.24$\,s/yr. We find a disagreeing value of $0.060 \pm 0.035$~s/yr, which would place in T~Vul rather in the third crossing of the instability strip \citepads{2014AstL...40..301F}.
Whether or not we include the interferometric data in the {\tt SPIPS} fit leads to consistent results, although the amplitude of the diameter variation tends to be slightly underestimated. 
We find a $p$-factor of $1.48 \pm 0.04_{\rm{stat}} \pm 0.18_{\rm{sys}}$, which is compatible at a level of 1$\sigma$ with most values deduced from published period-$p$ relations. In addition, our result agrees at 1.2$\sigma$ with the value of $1.19 \pm 0.16$ found in \citetads{2007AJ....133.1810B}. The final adjustment is shown in Fig.~\ref{TVul}.

\subsection{FF Aql}

\begin{figure*}
        \includegraphics[width=\hsize]{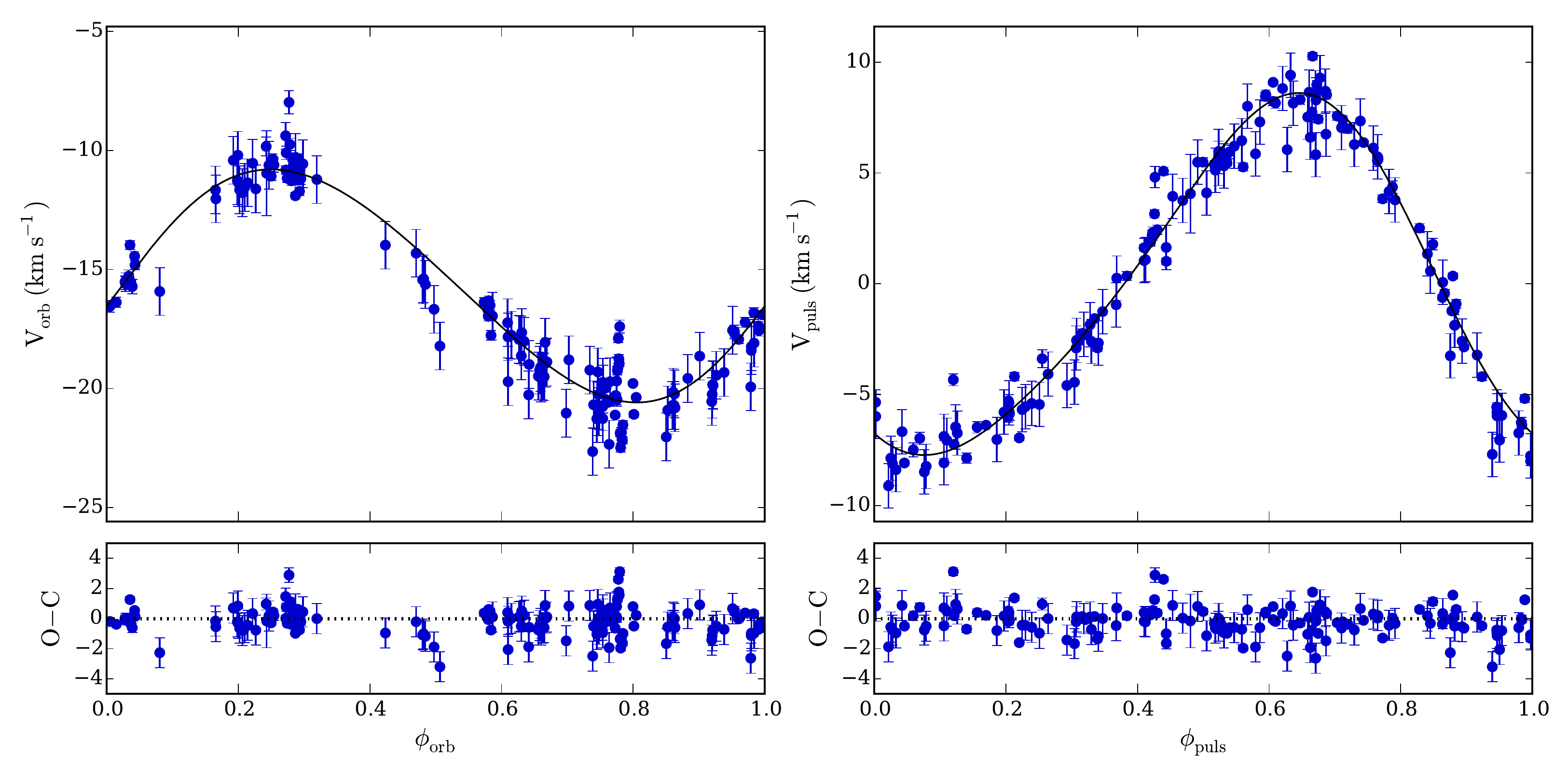}
        \caption{{\it Left:} Orbital velocity of FF Aql as a function of the orbital phase, corrected for the pulsation. {\it Right:} Pulsational radial velocity curve as a function of the pulsation phase.}
        \label{FFAql}
\end{figure*}

FF~Aql is known to be part of a possible quadruple system. The spectroscopic companion was recently studied with the VLT/NACO instrument, by \citetads{2014A&A...567A..60G}. No direct detection could be made, but the authors exclude spectral types outside from the A9V-F3V range.
The signature of orbital motion is clearly apparent in the RV curve, and was extensively studied by \citetads{1990AJ.....99.1598E}. 
We corrected all the data used in the present study \citepads{1990AJ.....99.1598E}, \citeads{1998AstL...24..815G} \citeads{1998MNRAS.297..825K}) with a modified version of the Wright \& Howard formalism \citepads{2009ApJS..182..205W}, in which we included the pulsation of the star (see \citeads{2013A&A...552A..21G}). We solved for the spectroscopic orbital elements and pulsation parameters with uncertainties derived via the bootstrapping technique (with replacement and 10000 bootstrap samples). Our derived parameters are an orbital period $P = 1438.76 \pm 1.60$~days, a JD of periastron passage $T = 2445204.81 \pm 87.17$~days, an eccentricity $e = 0.113 \pm 0.042$, an argument of periapsis $\omega = 262.7 \pm 23.6$, a velocity semi-amplitude $K = 4.949 \pm 0.156$~km\,s$^{-1}$, and a systemic velocity $V_{\gamma} = -15.6 \pm 0.1$~km\,s$^{-1}$. The reduced $\chi^{2}$ is 12.38 because of the relatively high intrinsic dispersion of the data. Both disentangled velocity curves are shown in Fig.~\ref{FFAql}. 
\citetads{2012A&A...541A..87G} published an average LD diameter of $0.878 \pm 0.013$~mas, consistent with our diameter of $0.870 \pm 0.013$~mas. Using the parallax from \citetads{2007AJ....133.1810B} ($\pi = 2.81 \pm 0.18$, $d = 356 \pm 23$~pc), we obtain a linear radius of $33.84 \pm 2.67 ~R_{\odot}$.
\citetads{2008MNRAS.389.1336K} and \citetads{2007AJ....133.1810B} published the same value for the color excess: $E(B-V)=0.224$, and \citetads{2013ApJ...772L..10T} give a similar value of $0.25 \pm 0.01$. We find a slightly lower reddening of $0.167 \pm 0.017$. \citetads{2013ApJ...772L..10T} also give an average temperature $\langle T_{\rm{eff}} \rangle = 6195 \pm 24$~K. Our temperature model is quite colder, but in agreement with the temperature published by \citetads{2011sf2a.conf..479G} ($\langle T_{\rm{eff}} \rangle = 5890 \pm 235$~K). \citetads{2013ApJ...772L..10T} place the Cepheid on the blue side of the instability strip and argue that the rate of period change ($+0.0703 \pm 0.0160$~s/yr) is consistent with this result.
\citetads{2014ARep...58..240B} also find a value of $dP/dt = 0.072 \pm 0.011$\,s/yr. We find that the Cepheid is close to the center of the instability strip with a very different rate of period change ($-0.140 \pm 0.036$\,s/yr), which places the Cepheid in the second crossing of the instability strip \citepads{2014AstL...40..301F}. We do not find any IR excess. A CSE has been brought out by \citetads{2011sf2a.conf..479G}, but it only becomes significant for $\lambda > 10~\mu m$.

The {\tt SPIPS} code for this particular Cepheid shows an irregular behavior. If we exclude the interferometric data, the amplitude of the diameter variation is highly underestimated and results in a much lower (and even unphysical) value of the $p$-factor: $p= 0.6 \pm 0.02_{\rm{stat}} \pm 0.07_{\rm{sys}}$. Including the interferometry in the fit leads to $p = 1.14 \pm 0.07_{\rm{stat}} \pm 0.07_{\rm{sys}}$, which is rather low (but still in a 3$\sigma$ agreement with most values deduced from the literature owing to the relatively large uncertainty). This could be due to a misestimation of the distance, which is actually subject to controversy. In particular, there is a tension between {\it Hipparcos} and {\it HST} parallaxes (yielding  to $d = 474 \pm 74$~pc and $d = 356 \pm 23$~pc, respectively). \citetads{2012A&A...543A..55N} also removed FF~Aql from their study of the $p$-factor because of this discrepancy, whose origin may be linked to the binary nature of the star. The relatively large uncertainties on the data could also explain a lower overall quality of the fitting process. The final $p$-factor value should therefore be considered with caution. The adjustment for FF~Aql is shown in Fig.~\ref{FFAql}.

\subsection{Y Sgr}


The 5.7-day period Cepheid Y~Sgr has not been as extensively studied as the rest of our sample. For this star, no interferometric observations are available, but the {\tt SPIPS} code nevertheless converged properly.
We corrected the different RV data sets to obtain average values and found the following offsets: $-3.59 \pm 0.04$~km\,s$^{-1}$ for \citetads{2002ApJS..140..465B}, $-1.79 \pm 0.02$~km\,s$^{-1}$ for \citetads{2009A&A...502..951N}, and $-2.55 \pm 0.03$~km\,s$^{-1}$ for \citetads{2011A&A...534A..94S}.
\citetads{1989CoKon..94.....S} underlines that the change in the $\gamma$-velocity could be due to the presence of a very long period ($>$10000 days) companion (he reports orbital variations in the O-C diagram).
\citetads{1992ApJ...384..220E} did not detect the companion in the data from the {\it International Ultraviolet Explorer}, but she set an upper limit on the spectral type, which could not be earlier than A2. \citetads{2002ApJS..140..465B} supports  the presence of a companion with an orbital period around 10000 days as well. There is no period change reported for this Cepheid.
Our study nevertheless leads to the value of $dP/dt = 0.016 \pm 0.048$~s/yr, which places Y~Sgr in the third crossing of the instability strip \citepads{2014AstL...40..301F}. \citetads{2008MNRAS.389.1336K} publish a color excess $E(B-V) = 0.182 \pm 0.021$, and \citetads{2007AJ....133.1810B} give the value of 0.205. Our own computation leads to $E(B-V) = 0.205 \pm 0.017$, which is in agreement with what we found in the literature. We find a linear radius of $43.10 \pm 6.73~R_{\odot}$, in agreement with the period-radius relationship from \citetads{2012MSAIS..19..205M}. Our effective temperature model is in average 400~K colder than the temperatures given by \citetads{2005AJ....130.1880A}. 
Our study leads to a $p$-factor of $p= 1.31 \pm 0.06_{\rm{stat}} \pm 0.18_{\rm{sys}}$, which is remarkably close to the values from \citetads{2013A&A...550A..70G}, \citetads{2007A&A...471..661N}, and \citetads{2012A&A...543A..55N}, and is compatible at a 1$\sigma$ level with most values deduced from the literature. The final adjustment for Y~Sgr is shown in Fig.~\ref{YSgr}.

\subsection{X Sgr}


X~Sgr is a 7-day period Cepheid known for the atypical features observed in its spectra, which are probably the consequence of the propagation of a double shockwave in the atmosphere of the star \citepads{2006A&A...457..575M}. Although this effect is expected to have an impact on the RV measurements, it does not seem to affect  our results much, since our output parameters are consistent with the average values found for the rest of the sample.
We determine a reddening $E(B-V)=0.286 \pm 0.016$, which is slightly higher than the values of $0.219$ found in \citetads{2008MNRAS.389.1336K}, and $0.197$ found in \citetads{2007AJ....133.1810B}. 
\citetads{2013A&A...549A..64L} determine a LD diameter of $1.48 \pm 0.08$~mas and a radius of $53 \pm 3~R_{\odot}$, considering the same distance as us (from \citeads{2007AJ....133.1810B}; $\pi = 3.00 \pm 0.18$, yielding to $d = 333 \pm 20$~pc).
These results are in slight tension (but consistent) with ours, as we determine an average diameter of $1.315 \pm 0.025$~mas, leading to a slightly lower radius of $48.37 \pm 3.84~R_{\odot}$.
\citetads{2008MNRAS.389.1336K} find a stellar diameter of $\theta_{\rm{LD}}=1.24 \pm 0.14$\,mas, and \citetads{2004A&A...416..941K} deduce from their VLTI/VINCI data a diameter $\theta_{\rm{UD}} = 1.471 \pm 0.033$~mas. Our {\tt SPIPS} analysis reveals significant excesses of 0.060 and 0.034~magnitude  in $K$ and $H$ bands, respectively.
A CSE around this Cepheid has been detected as a result of VLTI/MIDI observations by \citetads{2013A&A...558A.140G}, who found an excess of 7\% at $10.5~\mu m$. The same authors lead a detailed study of the envelope owing to the radiative transfer simulation code {\tt DUSTY}.
They find an average effective temperature for the star of $T_{\rm{eff}} = 5900$\,K, which is slightly lower than our value of $T_{\rm{eff}} = 6117 \pm 52$. These authors also find a lower reddening of $0.200 \pm 0.032$, which is actually closer to the value of \citetads{2008MNRAS.389.1336K}. The exclusion of the interferometry in the global fit leads to similar results, however, a slightly lower (although consistent) $p$-factor. This small instability could be due to the lack of RVs measurements, in particular, at the extrema. We find a linear period variation of $0.371 \pm 0.098$~s/yr, corresponding for X~Sgr to the third crossing of the instability strip \citepads{2014AstL...40..301F}. \citetads{1989CoKon..94.....S} finds a higher value of $0.74 \pm 0.09$~s/yr, which nevertheless corresponds to the same evolutionary status. 
Our study leads to a $p$-factor of $p= 1.39 \pm 0.04_{\rm{stat}} \pm 0.08_{\rm{sys}}$, which is remarkably similar to the value published by \citetads{2011A&A...534A..94S}, and agrees with most values deduced from the literature at a level of 1$\sigma$. The {\tt SPIPS} adjustment for this Cepheid is shown in Fig.~\ref{XSgr}.

\subsection{W Sgr}


W~Sgr is a 7.5-day Cepheid, which is known to belong to a triple system composed of a spectroscopic binary and a visual hot companion. The RVs from \citetads{1994A&AS..108...25B} are corrected for the orbital motion and perfectly match the data from \citetads{2005MNRAS.362.1167P}. The orbital elements were deduced again from {\it HST} observations of \citetads{2007AJ....133.1810B}, who found a period of 1582~days.
From \citetads{2009AJ....137.3700E}, this close companion (only 5~AU from the Cepheid) could not be earlier than an F0V star. The hot component located at 0.16~as could not be detected in NACO observations \citepads{2014A&A...567A..60G}. The study of the O-C diagram from \citetads{1989CoKon..94.....S} does not reveal a period variation. \citetads{1998JAVSO..26..101T} publishes a very low value of $-1.5$\,s/yr, which corresponds to the second crossing of the instability strip.
We find a very different value of $0.331 \pm 0.111$~s/yr, which places the Cepheid in the third crossing of the instability strip \citepads{2014AstL...40..301F}.

The VLTI/VINCI observations from \citetads{2004A&A...416..941K} lead to an average $\theta_{UD} = 1.312 \pm 0.029$~mas. We find a significantly lower diameter of $1.110 \pm 0.017$~mas, in agreement with \citetads{2011sf2a.conf..479G}.
Our PIONIER observations are undertaken in $H$ band, while the VINCI observations were made in the $K$ band. The star shows a large IR excess of about 0.106 magnitudes in $K$ and 0.064 magnitude in $H$. Its larger apparent size in the $K$ band is likely caused by the contribution of its extended CSE in this band. The code {\tt SPIPS} takes  the IR excess into account to fit all the interferometric data together.
The CSE was also brought out by \citetads{2011sf2a.conf..479G}, who detected a spatially resolved emission around W~Sgr in VLT/VISIR images. \citetads{2008MNRAS.389.1336K} and \citetads{2007AJ....133.1810B} give reddenings of  $E(B-V) = 0.079 \pm 0.017$ and $0.111 $, respectively. Our value stands in the middle, at $0.029 \pm 0.017$. 
Our study leads to a $p$-factor of $p= 1.35 \pm 0.06_{\rm{stat}} \pm 0.12_{\rm{sys}}$. This result is in agreement with most values deduced from the literature at a 1$\sigma$ level. In particular, it is remarkably close to the value found by \citetads{2012A&A...541A.134N}. The final adjustment for W~Sgr is shown in Fig.~\ref{WSgr}.

\subsection{$\beta$ Dor}


$\beta$~Dor is one of the brightest and biggest southern Cepheids, and it has therefore been extensively observed. Unlike most Cepheids, it has no visual or spectroscopic companion known, and we did not find any companion in our interferometric data. In order to reduce the dispersion of the RV curve of $\beta$~Dor, we corrected the three different RV data sets from their mean velocity (calculated from the model) to equal  0~km\,s$^{-1}$.
We found the following offsets: $8.63 \pm 0.03$~km\,s$^{-1}$ for \citetads{2002ApJS..140..465B}, $9.58 \pm 0.12$~km\,s$^{-1}$ for \citetads{2005MNRAS.362.1167P}, and $8.76 \pm 0.04$~km\,s$^{-1}$ for \citetads{2009A&A...502..951N}. We also decided to exclude the PIONIER measurement at $\phi = 0.78$, whose quality was low because of bad weather conditions during the observations.
Removing this point allows a higher stability of the fit. We observe that whether or not we use the interferometric data leads to the same final results, which confirms the robustness of the fitting process. The {\tt SPIPS} best-fit parameters for $\beta$ Dor give a reddening $E(B-V)= -0.018 \pm 0.016$, consistent with the value of 0.00 published by \citetads{2008MNRAS.389.1336K}.
The negative value could suggest the presence of an undetected hot companion. Our linear diameter is smaller than that published by \citetads{1998MNRAS.298..594T} ($R = 67.8 \pm 0.7~R_{\odot}$). However, they suggest a higher distance ($349 \pm 4$~pc) than the distance we use (from \citeads{2007AJ....133.1810B}; $\pi = 3.14 \pm 0.16$, or $d = 318 \pm 16$~pc), which makes both results consistent in terms of AD. \citetads{2004A&A...416..941K} found a value of $\theta_{UD} = 1.891 \pm 0.024$~mas, which is larger than our diameter of $1.776 \pm 0.012$~mas.
The O-C diagram from \citetads{1989CoKon..94.....S} does not suggest any period change. A more recent study led by the {\it Secret Lives of Cepheids} program \citepads{2015arXiv150402713E} finds a period change of $0.468 \pm 0.016$\,s/yr. In the present study, we find a value of $-0.084 \pm 0.149$\,s/yr, suggesting that $\beta$~Dor is in the second crossing of the instability strip \citepads{2014AstL...40..301F}. \citetads{1998JAVSO..26..101T} finds a much lower value of $-3.4$~s/yr.
We find an average effective temperature of $5318 \pm 51$~K, slightly lower than the value of $5490$ found in \citetads{2004A&A...416..941K}. 
Our study leads to a $p$-factor of $p= 1.36 \pm 0.04_{\rm{stat}} \pm 0.07_{\rm{sys}}$. This result is remarkably close to the values deduced from the period-$p$ relations published by \citetads{2012A&A...541A.134N} and \citetads{2011A&A...534A..94S}, and agrees with most other published values at a level of 1$\sigma$. The final adjustment for $\beta$~Dor is shown in Fig.~\ref{betaDor} of  Appendix A.

\subsection{$\zeta$ Gem}


As $\zeta$ Gem is the Northern Cepheid with the largest AD,  it has been the subject of a lot studies. This bright winter star is known to have a visual companion at 87\,arcseconds \citepads{1981A&AS...44..179P}, although it is uncertain if the stars are gravitationally bound.
We did not identify any close companion in our PIONIER data. However, we observe a slight variation of $V_{\rm{mean}}$ between the different data sets of RV that we used, which could be an actual variation of $V_{\gamma}$ due to orbital motion.
We determined and subtracted the following offsets: \citetads{1994A&AS..108...25B}, $5.939 \pm 0.085$~km\,s$^{-1}$; \citetads{1998AstL...24..815G}, $5.512 \pm 0.031$~km\,s$^{-1}$; \citetads{1998MNRAS.297..825K}, $6.603 \pm 0.136$~km\,s$^{-1}$; \citetads{2009A&A...502..951N}, $6.477 \pm 0.046$~km\,s$^{-1}$; and \citetads{2011A&A...534A..94S}, $7.363 \pm 0.017$~km\,s$^{-1}$.
However, such a small amplitude of variation (about 2~km\,s$^{-1}$) does not allow us to reach a conclusion about binarity, since it could also be due to instrumental systematics.
We find a negative but close to zero reddening, consistent with the values from \citetads{2008MNRAS.389.1336K} ($E(B-V) = 0.031 \pm 0.041$), \citetads{2007AJ....133.1810B} ($0.017$), and \citetads{2012ApJ...748L...9M} ($0.019 \pm 0.017$). \citetads{2012ApJ...748L...9M} established the membership of $\zeta$\,Gem to a host cluster lying at a distance $d=355 \pm 15$~pc, which is consistent with the distance used in the present study (from \citeads{2007AJ....133.1810B}; $\pi = 2.78 \pm 0.18$, $d = 360 \pm 23$\,pc).
We find a linear period variation of $-1.238 \pm 0.144$~s/yr, placing $\zeta$~Gem in the second crossing of the instability strip \citepads{2014AstL...40..301F}. \citetads{2015arXiv150402713E} propose a value of $-3.100 \pm 0.011$~s/yr, suggesting that the Cepheid could be either in its second or fourth crossing. Our whole temperature model is shifted by about 150~K compared to the $T_{\rm{eff}}$ measurements found in \citetads{2008AJ....136...98L}. From their VLTI/VINCI interferometric measurements, \citetads{2004A&A...416..941K} found an average diameter $\theta_{UD} = 1.747 \pm 0.061$~mas. This value is in agreement with the result of the present study ($1.663 \pm 0.049$~mas). 
Our study leads to a $p$-factor of $p= 1.41 \pm 0.04_{\rm{stat}} \pm 0.09_{\rm{sys}}$, which is in agreement at 1$\sigma$ with the values published by \citetads{2011A&A...534A..94S}, \citetads{2007A&A...471..661N}, and \citetads{2012A&A...541A.134N}, but is also compatible with most other published values at a 2$\sigma$ level. The final adjustment is shown in Fig.~\ref{zetaGem}.

\subsection{$\ell$ Car}


\begin{figure*} 
        \centering
        \resizebox{17cm}{!}{\includegraphics{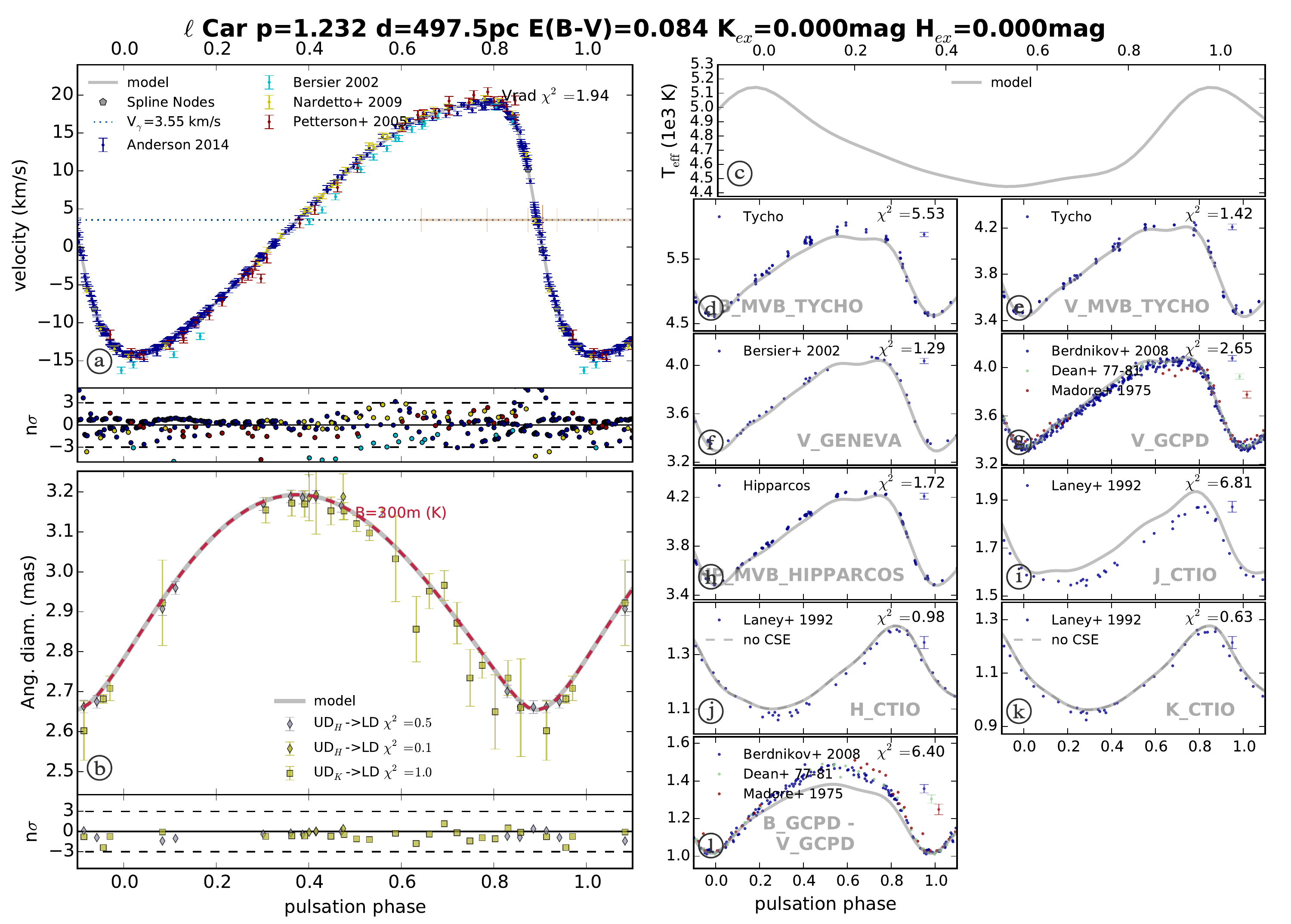}}
        \caption{Result of the SPIPS model fit to the observations of $\ell$ Car}
        \label{lCar}
\end{figure*}

\begin{figure}
        \resizebox{\hsize}{!}{\includegraphics{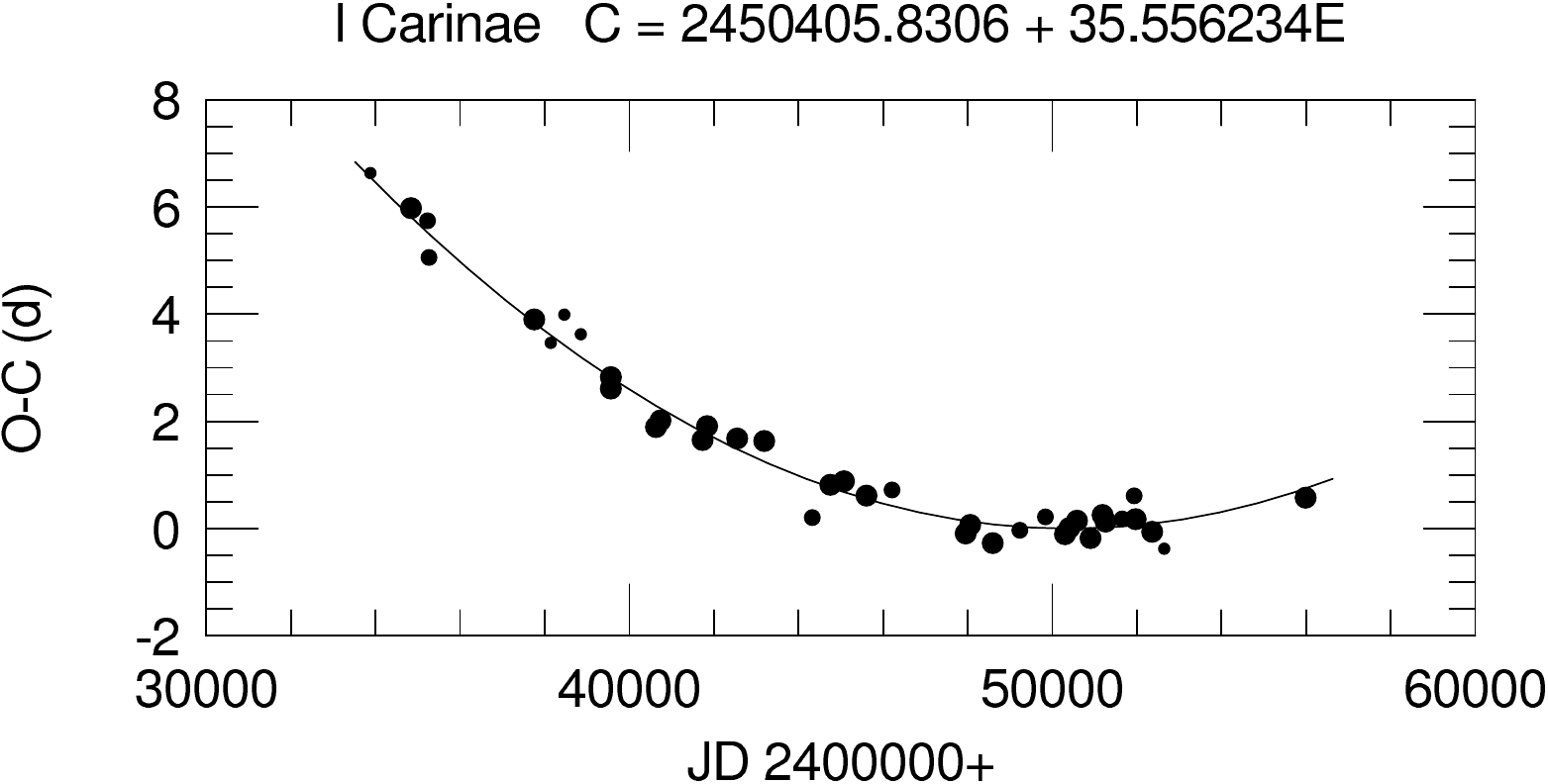}}
        \caption{O-C diagram of $\ell$\,Car.}
        \label{OCdiagramLCar}
\end{figure}

As discussed by \citetads{2014A&A...566L..10A}, the RV variations of $\ell$\,Car are not perfectly reproduced cycle-to-cycle. This is potentially a difficulty for the application of the BW technique, which relies on observational data sets that are generally obtained at different epochs and therefore different pulsation cycles. This potentially induces an uncertainty on the amplitude of the linear radius variation and therefore on the derived parameters (distance or $p$-factor).
However, as shown in Fig.~\ref{lCar}, the residual of the adjustment of the {\tt SPIPS} model is satisfactory in terms of RVs.
The quality of the fit is generally also good for the different photometric bands and colors for $\ell$\,Car. There is, however, a noticeable difference in the model predictions with the photometry for $\ell$\,Car in the deflation phase up to the minimum diameter rebound. This is an interesting feature, which is probably caused by a deviation of the surface brightness of $\ell$\,Car from the model atmosphere used in the {\tt SPIPS} code.
The interferometric ADs of $\ell$\,Car are accurately reproduced by the model, but a systematic shift of 3.5\% is present between the VINCI ($K$ band) and SUSI ($700$\,nm) measurements. The two PIONIER measurements ($H$ band) obtained shortly after the maximum radius phase are between the VINCI and SUSI. This may be interpreted as a bias due to the chosen LD model \citepads{2013A&A...554A..98N}.
However, the irregularity of the RV curve reported recently by \citetads{2014A&A...566L..10A} appears to be another possible reason for this effect, as the VINCI (epoch 2003) and SUSI (epoch 2004-2007) data were obtained during different pulsation cycles.
The O-C diagram of $\ell$\,Car is presented in Fig.~\ref{OCdiagramLCar}. The parabola fits the O-C residuals very well, and the minor fluctuations reflect the uncertainties in determining the moment of brightness maxima from sparsely covered light curves and the contribution of the intrinsic period noise present in Cepheids.
The secular increase in the pulsation period derived from the O-C diagram is $0.06225 \pm 0.00214$\,day/century, equivalent to $53.78 \pm 1.85$\,s/yr. The resulting ephemeris for brightness maxima of $\ell$\,Car is, therefore, (expressed in Julian date)
\begin{equation*}
\begin{split}
 C = & \ \  2450405.8306  \pm 0.0344\\
        & + \Bigl(35.556234\pm .000373\Bigr) \times E\\
        & + \Bigl(3.030 \times 10^{-5} \pm .104 \times 10^{-5}\Bigr) \times E^2.
\end{split}
\end{equation*}
The SPIPS code leads to a lower value than the O-C diagram, $27.283 \pm 0.984$~s/yr, which is also much lower than the value proposed by \citetads{1998JAVSO..26..101T} (118.5 s/yr).

The possible presence of an excess emission in the IR $H$ and $K$ bands is considered in the {\tt SPIPS} code, but no significant excess is detected in the present study. This is in contradiction with the detection reported in the $K$ band by \citetads{2006A&A...448..623K} for $\ell$\,Car. However, \citetads{2009A&A...498..425K} did not confirm the presence of a photometric excess in the $K$ band, although a considerable excess flux is found in the thermal IR ($10\,\mu$m) and longward.
The detection in the $K$ band reported by \citetads{2006A&A...448..623K} is based on the difference in visibility between observations of $\ell$\,Car that were obtained at short and long baselines. The detection of this excess relies implicitly on the assumption that the radial pulsation of the star repeats itself with an accuracy on the order of 1\%, i.e., that the stellar radius at a given phase is constant for different cycles.
If this is not the case, as argued by \citetads{2014A&A...566L..10A}, then the random difference in angular size can mimic the presence (or absence) of an envelope if the observations with the short and long baselines are not obtained within the same cycle, which was the case for the observations of \citetads{2006A&A...448..623K}.
For this same reason, based on our {\tt SPIPS} model, we do not exclude the presence of a CSE at a level of a few percent in the $K$ band.
Our study leads to a $p$-factor of $p= 1.23 \pm 0.01_{\rm{stat}} \pm 0.12_{\rm{sys}}$, which is in agreement with most results deduced from published period-$p$ relations.

\section{Discussion \label{Discussion}}

The $p$-factors resulting from the present study, and the main values deduced from published period-$p$ relations, are summarized in Table~\ref{pfactor_values}.

\begin{table*}
    \caption{$p$-factors calculated with {\tt SPIPS} and main values deduced from published period-$p$ relations.}
    \label{pfactor_values}
    \centering
    \renewcommand{\arraystretch}{1.4}
    \begin{tabular}{*{8}{c}}
        \hline
        Star & {\it present work} & $(1)$ & $(2)$ & $(3)$ & $(4)$ & $(5)$ & $(6)$ \\ 
        \hline
        \hline
RT Aur & 1.20 $\pm$ 0.12 & 1.363 $\pm$ 0.029 & 1.444 $\pm$ 0.074 & 1.339 $\pm$ 0.034 & 1.264 $\pm$ 0.089 & 1.356 $\pm$ 0.064 & 1.377 $\pm$ 0.003 \\
T Vul & 1.48 $\pm$ 0.18 & 1.345 $\pm$ 0.032 & 1.43 $\pm$ 0.079 & 1.335 $\pm$ 0.036 & 1.258 $\pm$ 0.092 & 1.344 $\pm$ 0.064 & 1.374 $\pm$ 0.003 \\
FF Aql & 1.14 $\pm$ 0.10 & 1.344 $\pm$ 0.033 & 1.429 $\pm$ 0.079 & 1.334 $\pm$ 0.036 & 1.258 $\pm$ 0.093 & 1.344 $\pm$ 0.064 & 1.373 $\pm$ 0.003 \\
Y Sgr & 1.31 $\pm$ 0.19 & 1.317 $\pm$ 0.038 & 1.408 $\pm$ 0.086 & 1.327 $\pm$ 0.038 & 1.249 $\pm$ 0.098 & 1.326 $\pm$ 0.064 & 1.368 $\pm$ 0.003 \\
X Sgr & 1.39 $\pm$ 0.09 & 1.297 $\pm$ 0.042 & 1.393 $\pm$ 0.091 & 1.322 $\pm$ 0.04 & 1.242 $\pm$ 0.102 & 1.313 $\pm$ 0.064 & 1.365 $\pm$ 0.003 \\
W Sgr & 1.35 $\pm$ 0.13 & 1.289 $\pm$ 0.044 & 1.386 $\pm$ 0.093 & 1.320 $\pm$ 0.041 & 1.240 $\pm$ 0.104 & 1.307 $\pm$ 0.064 & 1.363 $\pm$ 0.003 \\
$\beta$ Dor & 1.36 $\pm$ 0.08 & 1.262 $\pm$ 0.050 & 1.365 $\pm$ 0.100 & 1.312 $\pm$ 0.043 & 1.231 $\pm$ 0.110 & 1.289 $\pm$ 0.064 & 1.358 $\pm$ 0.003 \\
$\zeta$ Gem & 1.41 $\pm$ 0.10 & 1.258 $\pm$ 0.050 & 1.363 $\pm$ 0.100 & 1.312 $\pm$ 0.043 & 1.229 $\pm$ 0.110 & 1.287 $\pm$ 0.064 & 1.358 $\pm$ 0.003 \\
$\ell$ Car & 1.23 $\pm$ 0.12 & 1.128 $\pm$ 0.078 & 1.262 $\pm$ 0.133 & 1.277 $\pm$ 0.054 & 1.186 $\pm$ 0.138 & 1.200 $\pm$ 0.064 & 1.334 $\pm$ 0.003 \\
        \hline
    \end{tabular}
        \tablefoot{{\it References:} $(1)$ \citetads{2013A&A...550A..70G}; $(2)$ \citetads{2011A&A...534A..94S}; $(3)$ \citetads{2007A&A...471..661N}; $(4)$ \citetads{2009A&A...502..951N}; $(5)$ \citetads{2012A&A...543A..55N}; and $(6)$ \citetads{2012A&A...541A.134N}.}
\end{table*}



 

For almost all the Cepheids in the present study, the {\tt SPIPS} code converges toward the same $p$-factors whether or not we include  the interferometric data. This agreement is a confirmation that the surface brightness-color relations, which are implicitly included in the SPIPS atmosphere models, are reliable tools to determine ADs with photometry. This is an important asset to apply this technique to more distant Cepheids, both in our Galaxy and in nearby galaxies, for which interferometric measurements of their ADs are not feasible with the current instruments.
Our temperature models seem to be rather inconsistent with the spectroscopic $T_{\rm{eff}}$ values found in the literature (up to 400 kelvins of difference). However, both these measurements and our model have  significant error bars, which finally lead to statistical agreement.
The {\tt SPIPS} code also allows us to confirm the presence of bright CSEs for two Cepheids of our sample: X~Sgr and W~Sgr. They account respectively for $\approx 6$ and 10\% of the $K$ band flux of these stars. This is taken into account in the {\tt SPIPS} model estimation of the AD and photometry curves.

We added to our sample the prototype Cepheid $\delta$~Cep, whose $p$-factor has been measured with the HST/FGS parallax from \citetads{2002AJ....124.1695B} (\citeads{2005A&A...438L...9M}; see also \citeads{2015arXiv151001940M}).

The adjustment of a constant leads to the mean value of $p=1.306 \pm 0.027$ ($\chi^2 = 0.962$), while a linear regression gives a variation of $p=0.078_{\pm 0.123}(\log P-1)+1.316_{\pm 0.033}$ ($\chi^2 = 0.915$). 
An increase of $p$ with respect to the pulsation period is in contradiction with most (if not all) current results and predictions (see, for example, \citeads{2009A&A...502..951N, 2013A&A...550A..70G, 2012A&A...543A..55N, 2011A&A...534A..94S}).
However, we do not have a tight constrain on the slope because of the large uncertainties on the parallax values. We can therefore not reach a conclusion about an actual linear variation, but only suggest that our result is consistent with a constant $p$-factor within the uncertainties.



\begin{figure*}[htp]
  \centering
  \includegraphics[width=\hsize]{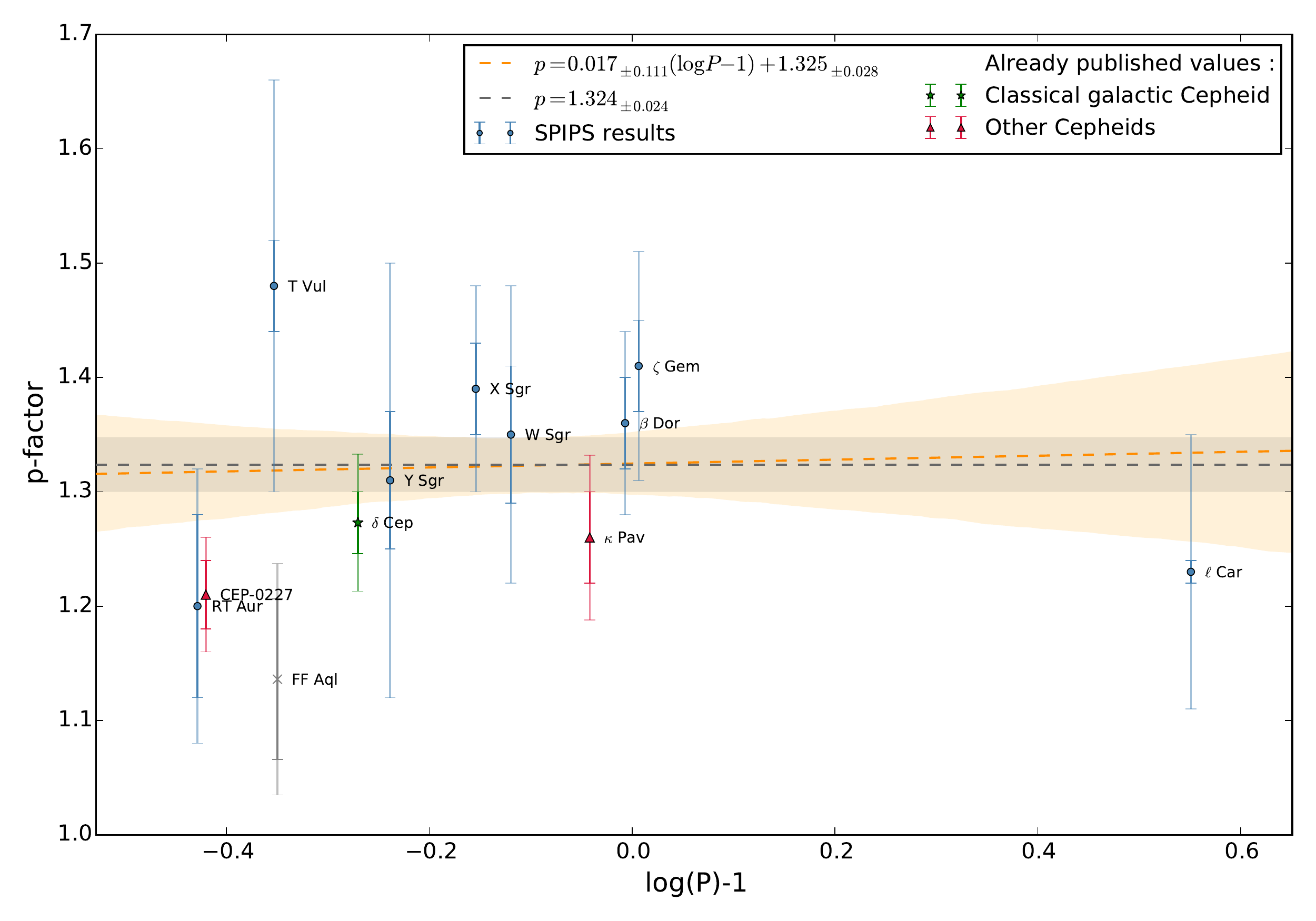}
  \caption{Relationship between the period and the $p$-factor for the Cepheids of our sample. We fitted both a constant (in black) and a linear regression (in orange). The error shades are defined by 1000 iterations of bootstrapping. The internal part of the error bar represents the statistical uncertainty. FF~Aql is plotted in gray because it has not been used in the final adjustments.}
  \label{pP}
\end{figure*}

As explained in Sect.~\ref{Results}, the {\tt SPIPS} code applied on FF~Aql shows an irregular behavior that makes us suspect a misestimation of the distance. We therefore decided to exclude it from the final adjustment. Figure~\ref{pP} shows the period-$p$ relation resulting from the nine other measurements. The new fit leads to a consistent average value of $p=1.324 \pm 0.024$ ($\chi^2 = 0.669$) and a shallower linear model of $p=0.017_{\pm 0.111}(\log P-1)+1.325_{\pm 0.028}$ ($\chi^2 = 0.667$).
Fig.~\ref{pP} also shows the $p$-factor values previously published for $\kappa$~Pav \citepads{2015A&A...576A..64B} and for the eclipsing binary Cepheid OGLE-LMC-CEP-0227 \citepads{2013MNRAS.436..953P}. Since the first  is a type II Cepheid and the second belongs to the Large Magellanic Cloud, they have lower metallicities and may exhibit slightly different properties. We therefore did not include them in the adjustment of the period-$p$ relation, although the results are not significantly different whether or not we consider them.

In the results presented above, all the uncertainties have been determined after 1000 iterations of bootstrapping, which implicitly averages the errors on the single $p$ measurements (i.e., on the {\it HST} parallaxes), although this is not justified because of the probable correlation between these errors. To adopt a more conservative approach, we therefore could use  the standard deviation of the residuals as a final
uncertainty, which would lead to the following results (when excluding FF~Aql from the adjustment): for the linear model, $p=0.017_{\pm 0.111}(\log P-1)+1.325_{\pm 0.085}$ and for the constant fit, $p=1.324 \pm 0.084$.

\section{Conclusion}

We presented {\tt SPIPS} models of the pulsation of nine Cepheids with available trigonometric parallaxes from \citetads{2007AJ....133.1810B}. We deduce the values of their spectroscopic $p$-factor, and their IR excess (the signature of the presence of a CSE) and their color excess $E(B-V)$ (caused by interstellar reddening).
Although the uncertainty of the parallaxes dominates the error bars on the derived $p$-factors, we conclude that within their total uncertainty, they are statistically consistent with a constant value independent of period, $p=1.324 \pm 0.024$.
The present calibration of the projection factor is limited by the relatively large uncertainty on the Cepheid parallaxes. As a result of the Gaia parallaxes that will be released over the next few years, we will soon be able to measure this essential parameter on a large sample of Galactic Cepheids with a sufficient accuracy to secure the PoP technique calibration at a 1\% level. 
Although precise distances will be known for a large number of galactic Cepheids in the Gaia era, the {\tt SPIPS} method will remain a very precious tool as it will lead to a better understanding of Cepheids physics (e.g., reddening, CSEs, etc.), essential for achieving the best precision and accuracy of the P-L relationships calibration. Thanks to the {\tt SPIPS} method, we will also be able to measure the distance of extragalactic Cepheids and study the dependance with metallicity. A simultaneous use of the Gaia data and the {\tt SPIPS} method will allow us to make a considerable step forward in the whole distance scale problematic.

\begin{acknowledgements}
This research received the support of PHASE, the partnership between ONERA, Observatoire de Paris, CNRS, and University Denis Diderot Paris 7.
AG acknowledges support from FONDECYT grant 3130361.
We acknowledge financial support from the ``Programme National de Physique Stellaire" (PNPS) of CNRS/INSU, France.
PK and AG acknowledge support of the French-Chilean exchange program ECOS-Sud/CONICYT.
We used the SIMBAD and VIZIER databases at the CDS, Strasbourg (France) and NASA's Astrophysics Data System.
This research has made use of the Jean-Marie Mariotti Center \texttt{SearchCal}, \texttt{LITpro,} and \texttt{Aspro}services (\url{http://www.jmmc.fr/}) codeveloped by FIZEAU and LAOG/IPAG.
\end{acknowledgements}

\bibliographystyle{aa} 
\bibliography{Benedict_language_edition.blg}

\begin{landscape}
\begin{table}
    \caption{Best-fit output parameters given by the \texttt{SPIPS} code.}
    \label{results}
    \centering
    \renewcommand{\arraystretch}{1.2}
    \begin{tabular}{*{10}{c}}
        \hline
        Name & Period (days) & \emph{p}-factor & $\theta_{\rm{Ross}~\rm{at}~\phi=0}$ (mas) & $E(B-V)$ & $<R>$ ($R_{\odot}$) & $<T_{\rm{eff}}>$ (K) & $V_{\gamma}$ (km/s) & $K$ excess (mag) & $H$ excess (mag) \B \T \\
        \hline
        \hline
\multirow{2}{*}{ RT Aur } & $\bf{ 3.728305 }$ & $\bf{ 1.20 }$ & $\bf{ 0.6772 }$ & $\bf{ 0.048 }$ & $\bf{ 31.55 }$ & $\bf{ 5876 }$ & $\bf{ 19.65 }$ & $\bf{ 0.000 }$ & $\bf{ 0.000 }$ \T \\
& \footnotesize{ 0.000005 }&  \footnotesize{$~\pm 0.08 ~\pm 0.09 $} & \footnotesize{ $~\pm 0.0025 ~\pm 0.0135 $} & \footnotesize{ $~\pm 0.008 ~\pm 0.016 $} & \footnotesize{ $~\pm 0.12 ~\pm 3.13 $} & \footnotesize{ $~\pm 23 ~\pm 50 $} & \footnotesize{ $~\pm 1.90 ~\pm 0.30 $} & \footnotesize{ $~\pm 0.006 ~\pm 0.020 $} & \footnotesize{ $~\pm 0.007 ~\pm 0.020 $} \B \\
\arrayrulecolor[gray]{.6}\hline
\multirow{2}{*}{ T Vul } & $\bf{ 4.435424 }$ & $\bf{ 1.48 }$ & $\bf{ 0.6071 }$ & $\bf{ 0.019 }$ & $\bf{ 35.39 }$ & $\bf{ 5736 }$ & \multirow{2}{*}{see section~\ref{Results}} & $\bf{ 0.004 }$ & $\bf{ 0.000 }$ \T \\
& \footnotesize{ 0.000005 }&  \footnotesize{$~\pm 0.04 ~\pm 0.18 $} & \footnotesize{ $~\pm 0.0012 ~\pm 0.0120 $} & \footnotesize{ $~\pm 0.005 ~\pm 0.016 $} & \footnotesize{ $~\pm 0.07 ~\pm 4.98 $} & \footnotesize{ $~\pm 13 ~\pm 50 $} &    & \footnotesize{ $~\pm 0.001 ~\pm 0.020 $} & \footnotesize{ $~\pm 0.004 ~\pm 0.020 $} \B \\
\arrayrulecolor[gray]{.6}\hline
\multirow{2}{*}{ FF Aql } & $\bf{ 4.470848 }$ & $\bf{ 1.14 }$ & $\bf{ 0.8703 }$ & $\bf{ 0.167 }$ & $\bf{ 33.84 }$ & $\bf{ 5823 }$ & \multirow{2}{*}{see section~\ref{Results}} & $\bf{ 0.000 }$ & $\bf{ 0.000 }$ \T \\
& \footnotesize{ 0.000010 }&  \footnotesize{$~\pm 0.07 ~\pm 0.07 $} & \footnotesize{ $~\pm 0.0014 ~\pm 0.0130 $} & \footnotesize{ $~\pm 0.007 ~\pm 0.016 $} & \footnotesize{ $~\pm 0.06 ~\pm 2.67 $} & \footnotesize{ $~\pm 21 ~\pm 50 $} &    & \footnotesize{ $~\pm 0.009 ~\pm 0.020 $} & \footnotesize{ $~\pm 0.009 ~\pm 0.020 $} \B \\
\arrayrulecolor[gray]{.6}\hline
\multirow{2}{*}{ Y Sgr } & $\bf{ 5.773383 }$ & $\bf{ 1.31 }$ & $\bf{ 0.8221 }$ & $\bf{ 0.205 }$ & $\bf{ 43.10 }$ & $\bf{ 5623 }$ & \multirow{2}{*}{see section~\ref{Results}} & $\bf{ 0.000 }$ & $\bf{ 0.011 }$ \T \\
& \footnotesize{ 0.000009 }&  \footnotesize{$~\pm 0.06 ~\pm 0.18 $} & \footnotesize{ $~\pm 0.0027 ~\pm 0.0164 $} & \footnotesize{ $~\pm 0.007 ~\pm 0.016 $} & \footnotesize{ $~\pm 0.14 ~\pm 6.73 $} & \footnotesize{ $~\pm 19 ~\pm 50 $} &    & \footnotesize{ $~\pm 0.009 ~\pm 0.020 $} & \footnotesize{ $~\pm 0.009 ~\pm 0.020 $} \B \\
\arrayrulecolor[gray]{.6}\hline
\multirow{2}{*}{ X Sgr } & $\bf{ 7.012770 }$ & $\bf{ 1.39 }$ & $\bf{ 1.3157 }$ & $\bf{ 0.286 }$ & $\bf{ 48.37 }$ & $\bf{ 6117 }$ & $\bf{ -13.15 }$ & $\bf{ 0.060 }$ & $\bf{ 0.034 }$ \T \\
& \footnotesize{ 0.000012 }&  \footnotesize{$~\pm 0.04 ~\pm 0.08 $} & \footnotesize{ $~\pm 0.0023 ~\pm 0.0255 $} & \footnotesize{ $~\pm 0.004 ~\pm 0.016 $} & \footnotesize{ $~\pm 0.08 ~\pm 3.84 $} & \footnotesize{ $~\pm 15 ~\pm 50 $} & \footnotesize{ $~\pm 0.07 ~\pm 0.30 $} & \footnotesize{ $~\pm 0.004 ~\pm 0.020 $} & \footnotesize{ $~\pm 0.004 ~\pm 0.020 $} \B \\
\arrayrulecolor[gray]{.6}\hline
\multirow{2}{*}{ W Sgr } & $\bf{ 7.594984 }$ & $\bf{ 1.35 }$ & $\bf{ 1.1106 }$ & $\bf{ 0.029 }$ & $\bf{ 54.60 }$ & $\bf{ 5497 }$ & $\bf{ -27.94 }$ & $\bf{ 0.106 }$ & $\bf{ 0.064 }$ \T \\
& \footnotesize{ 0.000009 }&  \footnotesize{$~\pm 0.06 ~\pm 0.12 $} & \footnotesize{ $~\pm 0.0034 ~\pm 0.0170 $} & \footnotesize{ $~\pm 0.008 ~\pm 0.016 $} & \footnotesize{ $~\pm 0.17 ~\pm 5.63 $} & \footnotesize{ $~\pm 20 ~\pm 50 $} & \footnotesize{ $~\pm 0.04 ~\pm 0.30 $} & \footnotesize{ $~\pm 0.005 ~\pm 0.020 $} & \footnotesize{ $~\pm 0.020 ~\pm 0.020 $} \B \\
\arrayrulecolor[gray]{.6}\hline
\multirow{2}{*}{ $\beta$ Dor } & $\bf{ 9.842675 }$ & $\bf{ 1.36 }$ & $\bf{ 1.7767 }$ & $\bf{ -0.018 }$ & $\bf{ 62.12 }$ & $\bf{ 5318 }$ & \multirow{2}{*}{see section~\ref{Results}} & $\bf{ 0.021 }$ & $\bf{ 0.000 }$ \T \\
& \footnotesize{ 0.000019 }&  \footnotesize{$~\pm 0.04 ~\pm 0.07 $} & \footnotesize{ $~\pm 0.0024 ~\pm 0.0120 $} & \footnotesize{ $~\pm 0.005 ~\pm 0.016 $} & \footnotesize{ $~\pm 0.09 ~\pm 3.58 $} & \footnotesize{ $~\pm 12 ~\pm 50 $} &    & \footnotesize{ $~\pm 0.006 ~\pm 0.020 $} & \footnotesize{ $~\pm 0.005 ~\pm 0.020 $} \B \\
\arrayrulecolor[gray]{.6}\hline
\multirow{2}{*}{ $\zeta$ Gem } & $\bf{ 10.149806 }$ & $\bf{ 1.41 }$ & $\bf{ 1.6633 }$ & $\bf{ -0.021 }$ & $\bf{ 64.87 }$ & $\bf{ 5326 }$ & \multirow{2}{*}{see section~\ref{Results}} & $\bf{ 0.000 }$ & $\bf{ 0.000 }$ \T \\
& \footnotesize{ 0.000017 }&  \footnotesize{$~\pm 0.04 ~\pm 0.09 $} & \footnotesize{ $~\pm 0.0019 ~\pm 0.0491 $} & \footnotesize{ $~\pm 0.006 ~\pm 0.016 $} & \footnotesize{ $~\pm 0.08 ~\pm 6.12 $} & \footnotesize{ $~\pm 16 ~\pm 50 $} &    & \footnotesize{ $~\pm 0.007 ~\pm 0.020 $} & \footnotesize{ $~\pm 0.007 ~\pm 0.020 $} \B \\
\arrayrulecolor[gray]{.6}\hline
\multirow{2}{*}{ $\ell$ Car } & $\bf{ 35.551609 }$ & $\bf{ 1.23 }$ & $\bf{ 2.7805 }$ & $\bf{ 0.084 }$ & $\bf{ 159.01 }$ & $\bf{ 4696 }$ & $\bf{ 3.55 }$ & $\bf{ 0.000 }$ & $\bf{ 0.000 }$ \T \\
& \footnotesize{ 0.000265 }&  \footnotesize{$~\pm 0.01 ~\pm 0.12 $} & \footnotesize{ $~\pm 0.0038 ~\pm 0.0283 $} & \footnotesize{ $~\pm 0.007 ~\pm 0.016 $} & \footnotesize{ $~\pm 0.21 ~\pm 17.44 $} & \footnotesize{ $~\pm 14 ~\pm 50 $} & \footnotesize{ $~\pm 0.04 ~\pm 0.30 $} & \footnotesize{ $~\pm 0.007 ~\pm 0.020 $} & \footnotesize{ $~\pm 0.005 ~\pm 0.020 $} \B \\
\arrayrulecolor[gray]{.6}\hline
    \end{tabular}
\end{table}
\end{landscape}

\include{appendix}

\end{document}

%% file: appendix.tex
\appendix
\onecolumn

\section{{\tt SPIPS} model for each Cepheid}

The figures in this Appendix show the result of the {\tt SPIPS} modeling of the stars of our sample (apart from $\ell$\,Car, that is presented in Fig.~\ref{lCar}). In all plots, the model is represented using a grey curve.

\begin{figure*}[h!]
	\centering
	\includegraphics[width=14cm]{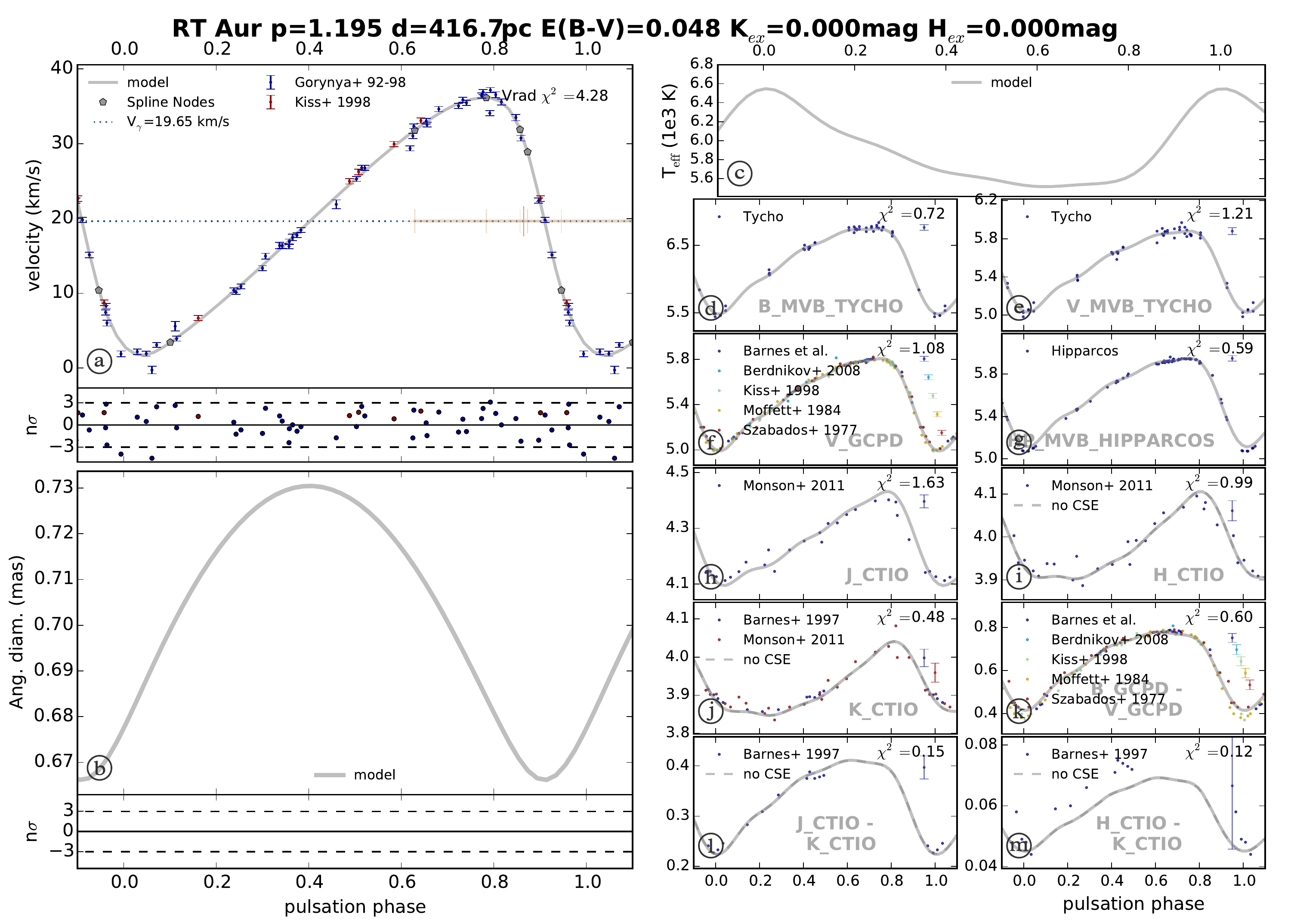}
	\caption{{\tt SPIPS} model of RT Aur.}
	\label{RTAur}
\end{figure*}

\begin{figure*}[h!]
	\centering
	\includegraphics[width=14cm]{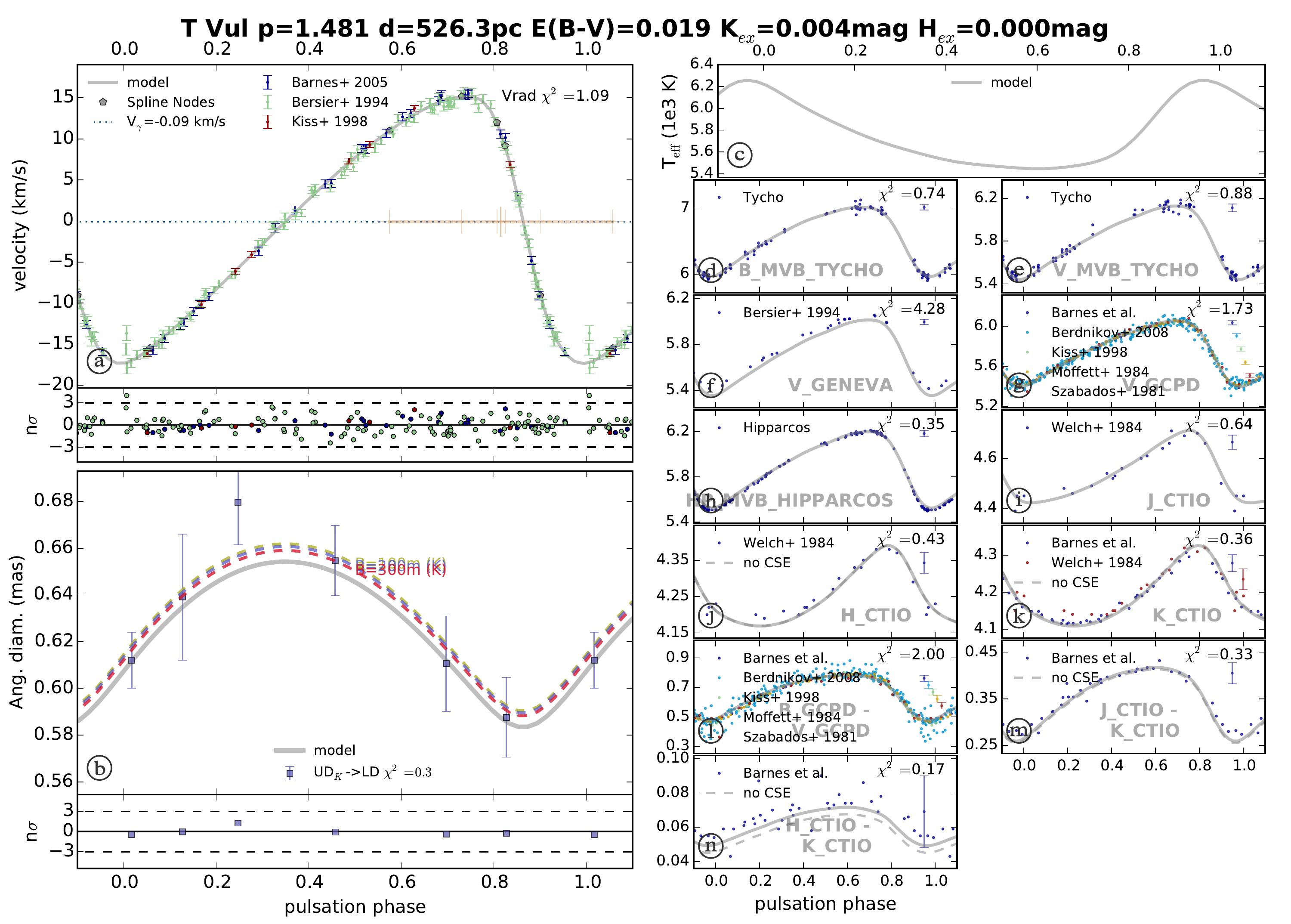}
	\caption{{\tt SPIPS} model of T Vul.}
	\label{TVul}
\end{figure*}

\begin{figure*}
	\centering
	\includegraphics[width=15cm]{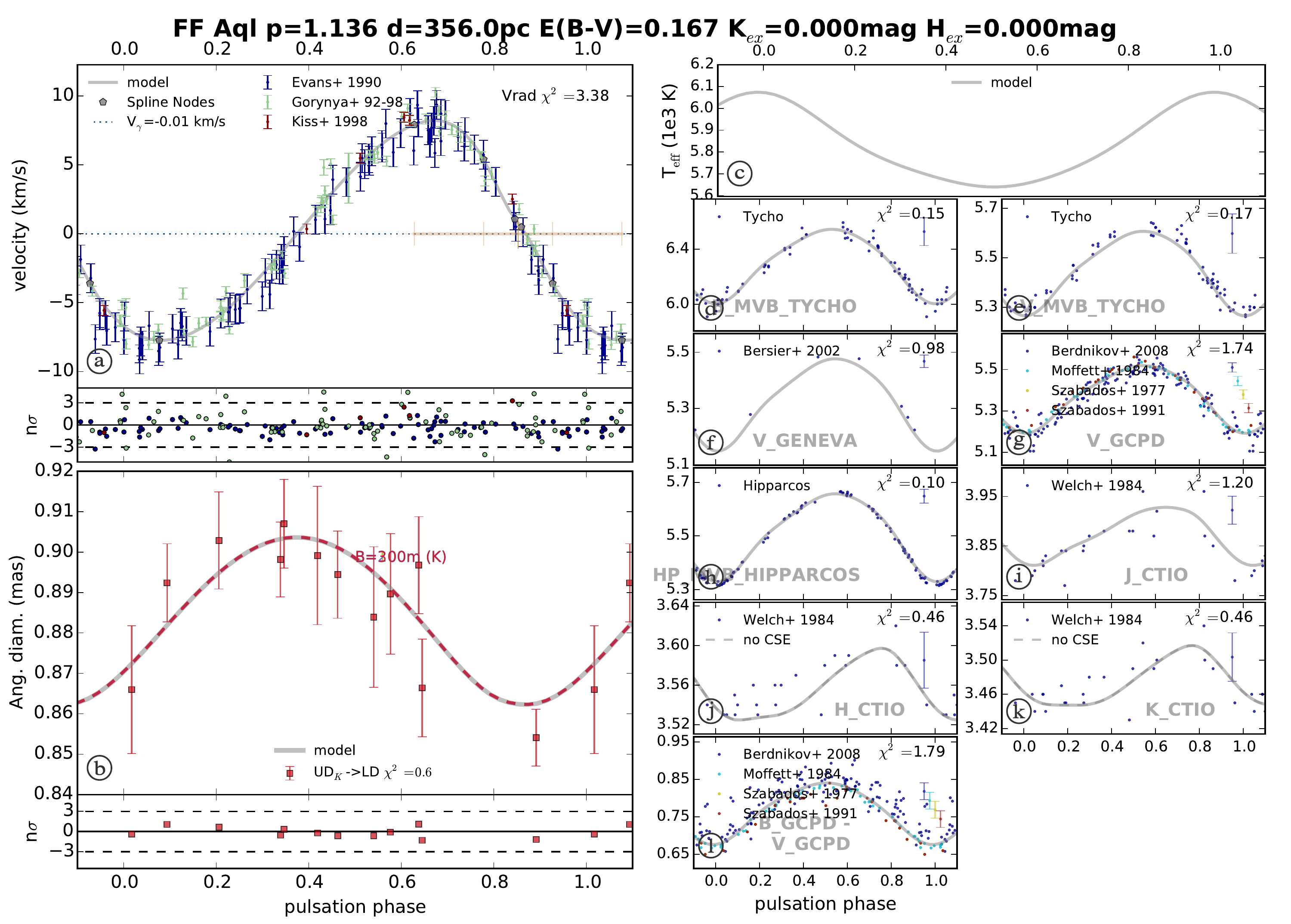}
	\caption{{\tt SPIPS} model of FF Aql.}
	\label{FFAql}
\end{figure*}

\begin{figure*}
	\centering
	\includegraphics[width=15cm]{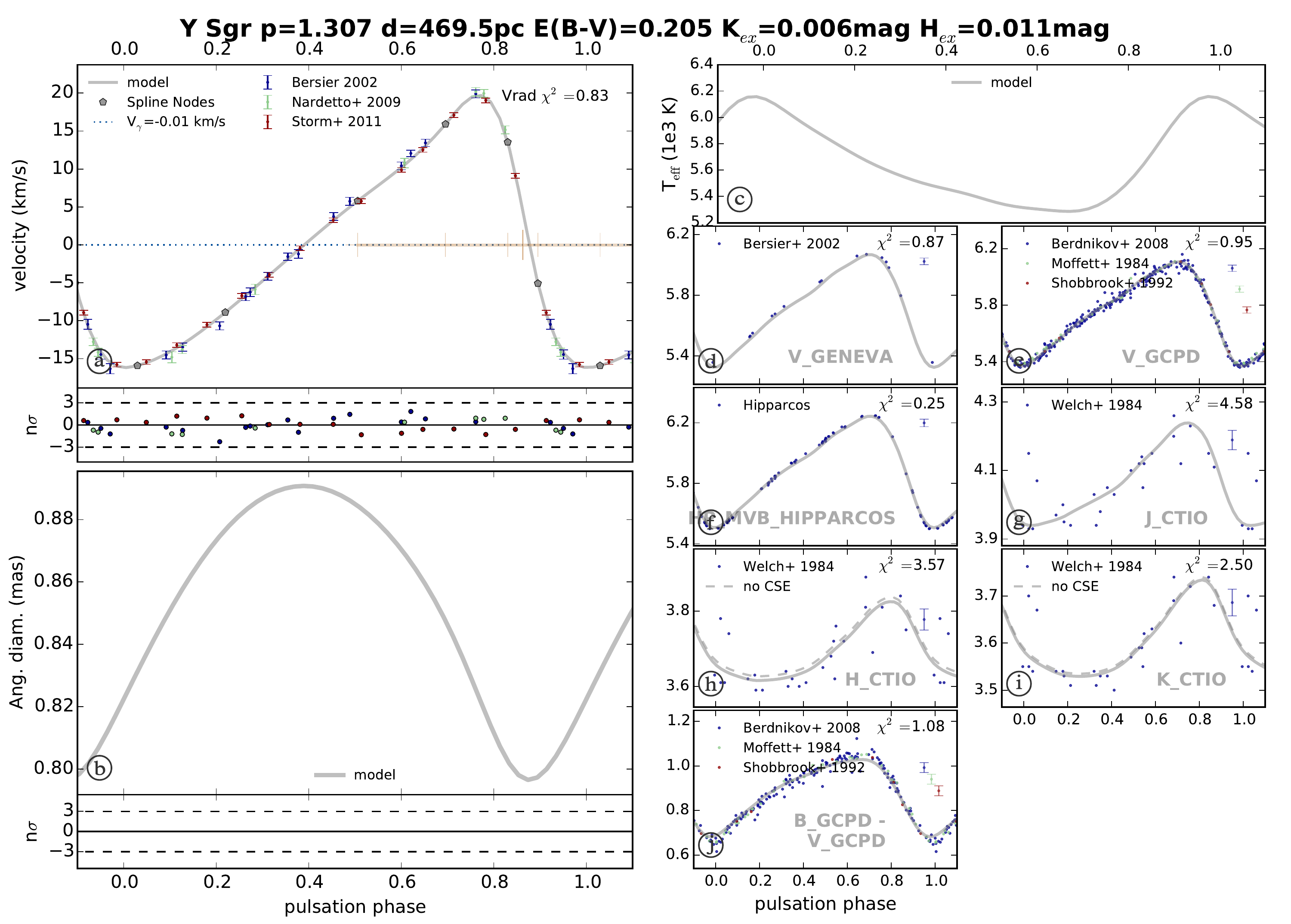}
	\caption{{\tt SPIPS} model of Y Sgr.}
	\label{YSgr}
\end{figure*}

\begin{figure*}
	\centering
	\includegraphics[width=15cm]{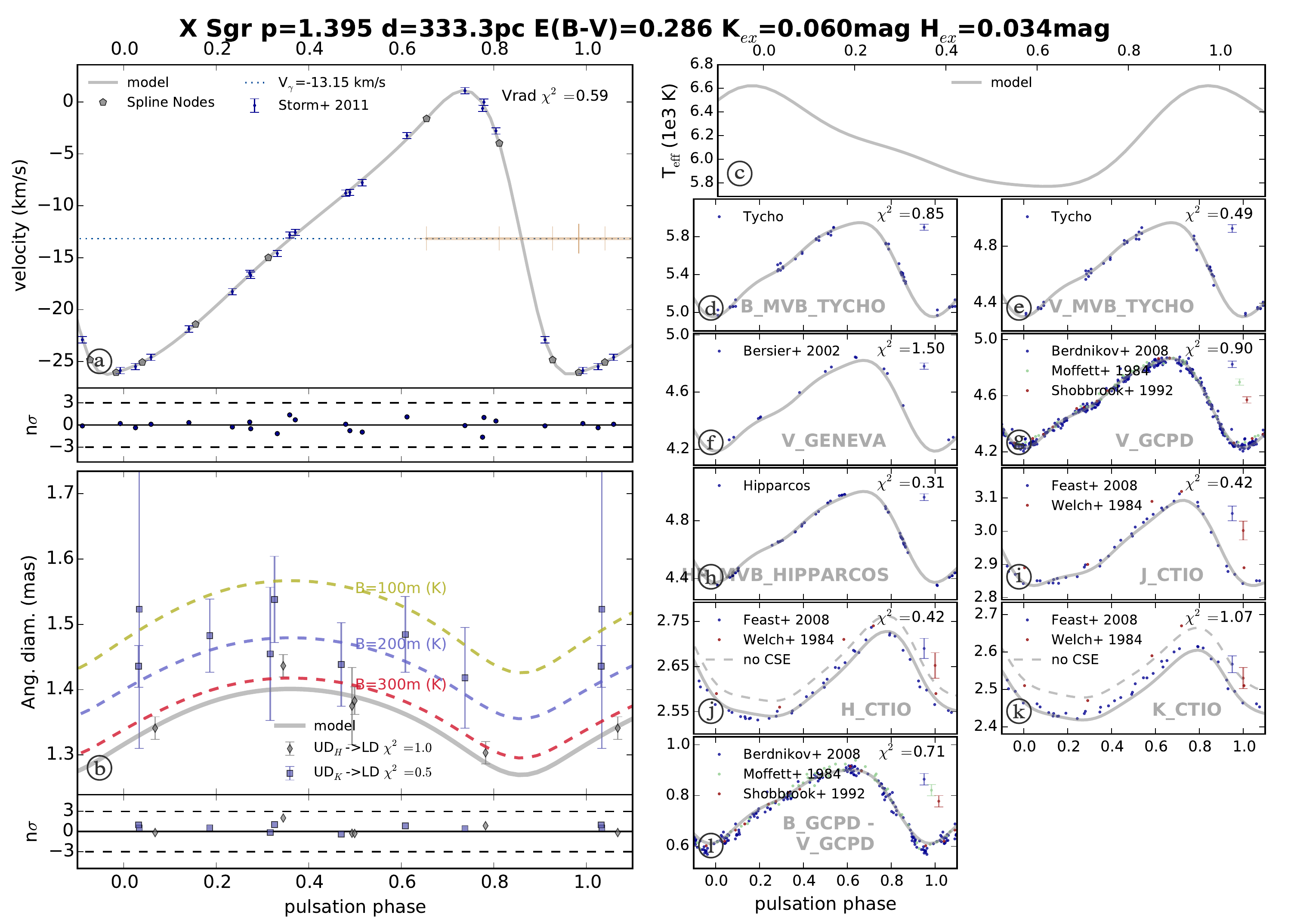}
	\caption{{\tt SPIPS} model of X Sgr.}
	\label{XSgr}
\end{figure*}

\begin{figure*}
	\centering
	\includegraphics[width=15cm]{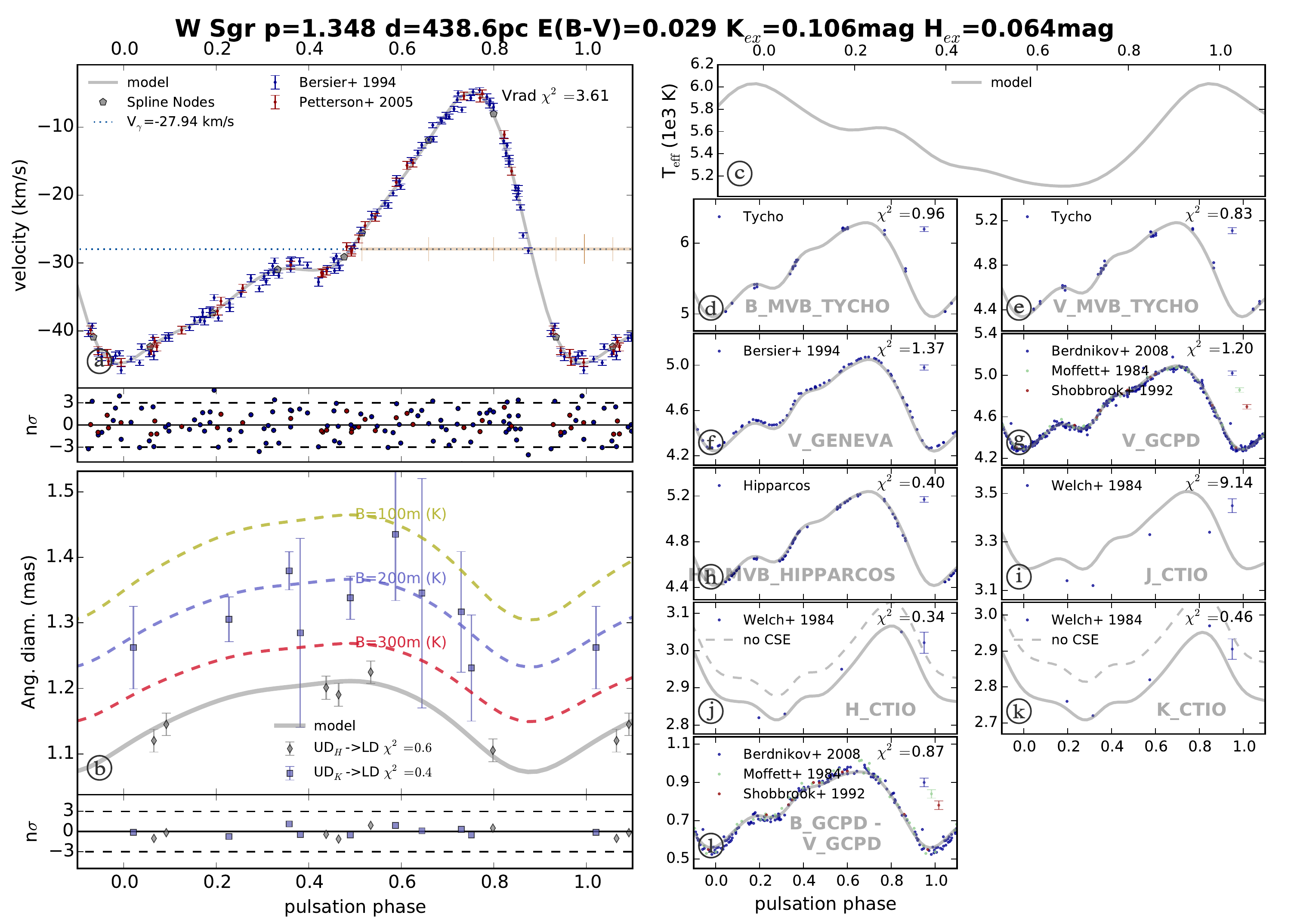}
	\caption{{\tt SPIPS} model of W Sgr.}
	\label{WSgr}
\end{figure*}

\begin{figure*}
	\centering
	\includegraphics[width=15cm]{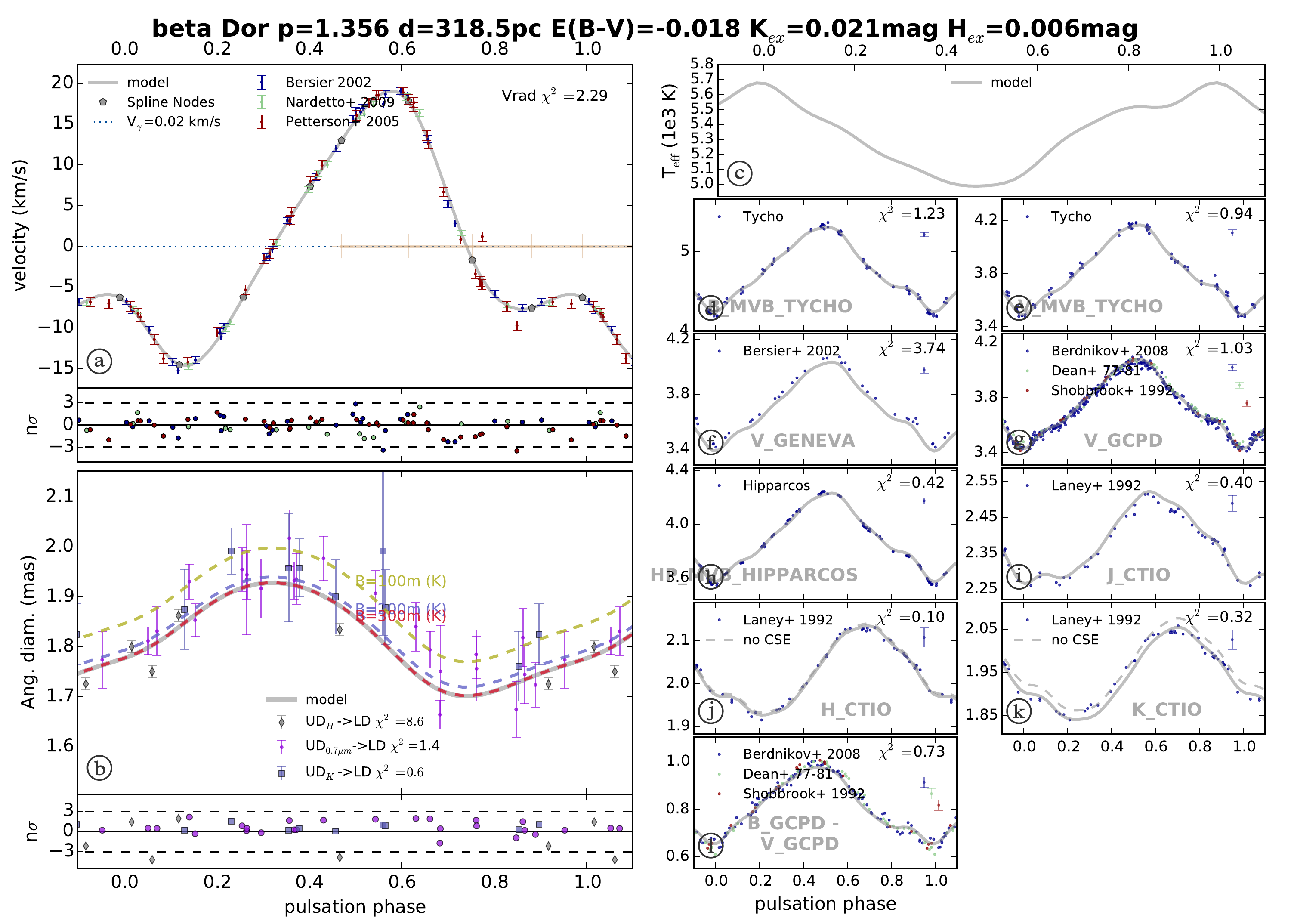}
	\caption{{\tt SPIPS} model of $\beta$ Dor.}
	\label{betaDor}
\end{figure*}

\begin{figure*}
	\centering
	\includegraphics[width=15cm]{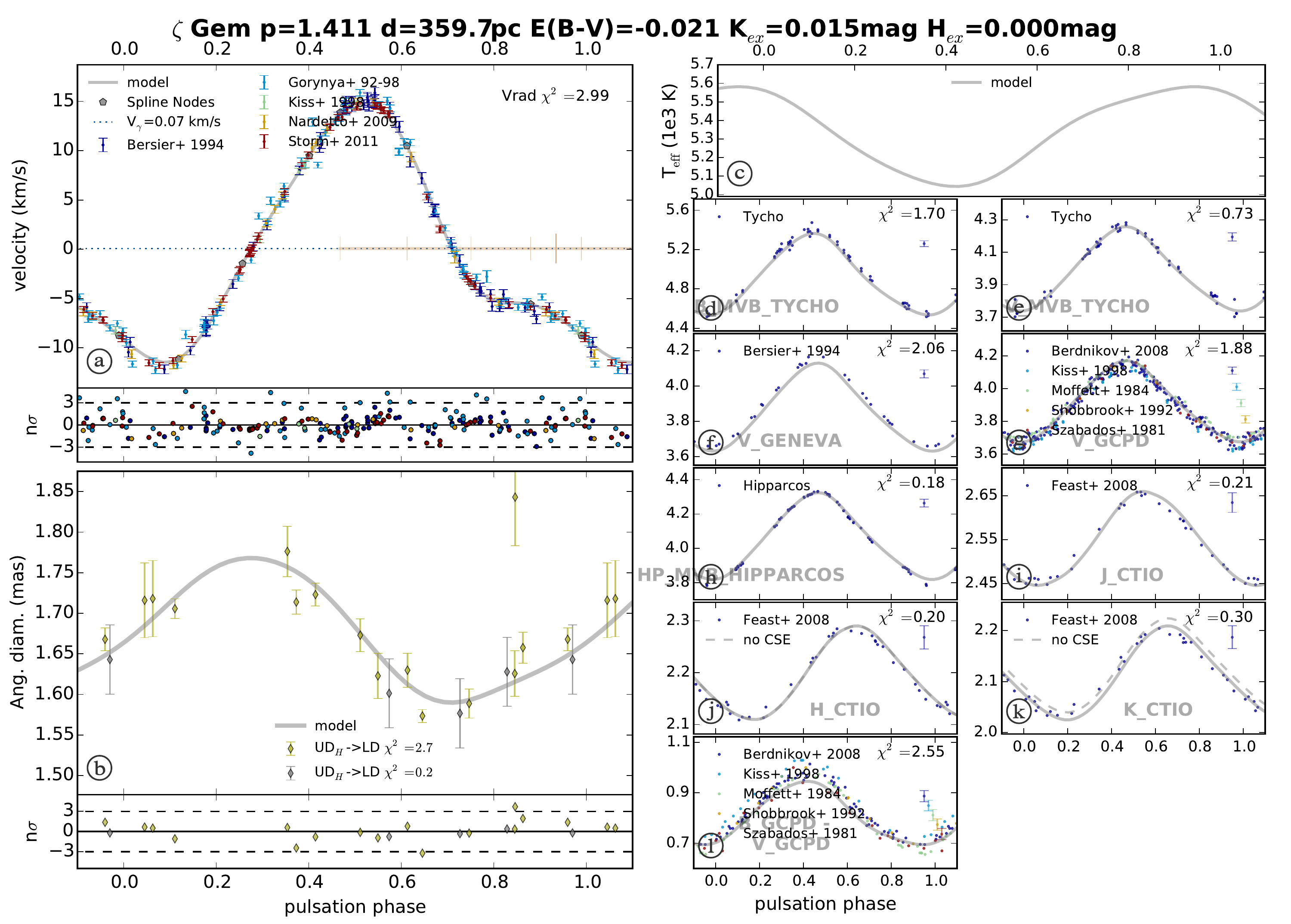}
	\caption{{\tt SPIPS} model of $\zeta$ Gem.}
	\label{zetaGem}
\end{figure*}